\documentclass[fleqn,usenatbib]{mnras}

\usepackage{newtxtext,newtxmath}

\usepackage[T1]{fontenc}

\DeclareRobustCommand{\VAN}[3]{#2}
\let\VANthebibliography\thebibliography
\def\thebibliography{\DeclareRobustCommand{\VAN}[3]{##3}\VANthebibliography}

\usepackage[T1]{fontenc}
\usepackage{Journal_names}
\usepackage{graphicx}
\usepackage{dcolumn}
\usepackage{bm}
\usepackage{amsmath}
\usepackage{xspace}
\usepackage{tikz}
\usetikzlibrary{fadings, shapes.geometric}
\usepackage{tabularx}
\usepackage{booktabs}
\usepackage{array}
\usepackage{fleqn}
\usepackage{float}
\usepackage{ragged2e}
\usepackage{subcaption}
\usepackage{multirow}
\makeatletter
\renewenvironment{figure}[1][tbp]
  {%
   \let\@makecaption\SFB@makefigurecaption
   \@float{figure}[#1]\fstyle@figure}
  {\end@float}
\makeatother

\makeatletter
\long\def\@makecaption#1#2{%
  \vskip\abovecaptionskip
  \centering 
  \parbox{\linewidth}{\justifying\small #1. #2\par}%
  \vskip\belowcaptionskip}
\makeatother

\newcolumntype{P}[1]{>{\centering\arraybackslash}p{#1}}
\usepackage[mathlines]{lineno}

\usepackage{graphicx}	
\usepackage{amsmath}	

\newcommand{\Rg}{\ensuremath{R_{\rm g}}}
\newcommand{\multistar}{{\tt multistar}\xspace}

\newcommand{\pc}{\ensuremath{\mathrm{\ pc}}}
\newcommand{\cl}{\ensuremath{\mathrm{c}}}
\newcommand{\G}{\ensuremath{\mathrm{G}}}

\newcommand{\dd}{\mathrm{d}}
\newcommand{\Gyr}{\mathrm{Gyr}}

\newcommand{\mMBH}{\ensuremath{M_{\rm BH}}}

\newcommand{\rpDM}{\ensuremath{\mathfrak{R}_{\rm p}}}

\newcommand{\rpDMz}{\ensuremath{\mathfrak{R}_{\rm p, 0}}}
\newcommand{\raDM}{\ensuremath{\mathfrak{R}_{\rm a}}}
\newcommand{\aDM}{\ensuremath{\mathcal{A}}}

\newcommand{\raDMz}{\ensuremath{\mathfrak{R}_{\rm a, 0}}}
\newcommand{\aEMRI}{\ensuremath{\mathfrak{a}}}

\newcommand{\aEMRIz}{\ensuremath{\mathfrak{a}_0}}

\newcommand{\aMBHB}{\ensuremath{\mathscr{a}}}

\newcommand{\aMBHBz}{\ensuremath{\mathscr{a}_0}}

\newcommand{\iotaz}{\ensuremath{\iota_0}}
\newcommand{\eDM}{\ensuremath{\mathcal{E}}}

\newcommand{\eDMz}{\ensuremath{\mathcal{E}}_0}

\newcommand{\Mp}{\ensuremath{M_1}}
\newcommand{\Ms}{\ensuremath{M_2}}
\newcommand{\Mps}{\ensuremath{M_{12}}}
\newcommand{\ME}{\ensuremath{M_{\rm s}}}
\newcommand{\Rschp}{\ensuremath{R_{\rm s, 1}}}
\newcommand{\Rschs}{\ensuremath{R_{\rm s, 2}}}
\newcommand{\Rschps}{\ensuremath{R_{\rm s, 12}}}
\newcommand{\Msolar}{\ensuremath{\mathrm{M}_{\odot}}}

\newcommand{\V}[1]{\mathbf{#1}}

\definecolor{cornellGreen}{HTML}{6EB43F}

\defcitealias{Gondolo_1999}{G\&S}
\newcommand{\GS}{\citetalias{Gondolo_1999}\xspace}

\title[Depletion of DM Spikes by MBH Mergers]{Depletion of Dark Matter Spikes by Massive Black Hole Mergers}

\author[Sharpe et al.]{Charlie Sharpe,$^{1}$\thanks{charles.sharpe@physics.ox.ac.uk}
Thomas F.~M.~Spieksma,$^{1,2}$\,
Yonadav Barry Ginat,$^{1}$\,
Ricarda S. Beckmann,$^{3}$\,\newauthor
Alexander Heger$^{4}$\,
and Bence Kocsis$^{1,5,6}$
\thanks{MTA Distinguished Guest Scientist Fellowship Programme 2026}
\\
$^{1}$Department of Physics, University of Oxford, Parks Road, Oxford, OX1 3PU, United Kingdom\\
$^{2}$School of Natural Sciences, Institute for Advanced Study, 1 Einstein Drive, Princeton, NJ 08540, USA\\
$^{3}$Institute for Astronomy, University of Edinburgh, Royal Observatory, Blackford Hill, Edinburgh, EH9 3HJ, United Kingdom\\
$^{4}$School of Physics and Astronomy, Monash University, 19 College Walk, Melbourne, Victoria 3800, Australia\\
$^{5}$St.~Hugh's College, University of Oxford, St.~Margaret's Road, Oxford, OX2 6LE, United Kingdom\\
$^{6}$Institute of Physics, Eötvös University, Pázmány P.~s.~1/A, Budapest 1117, Hungary}

\date{Accepted XXX. Received YYY; in original form ZZZ}

\pubyear{\the\year{}}

\begin{document}
\label{firstpage}
\pagerange{\pageref{firstpage}--\pageref{lastpage}}
\maketitle

\begin{abstract}
Dark matter over-densities around massive black holes, known as ``dark matter spikes'', can dephase gravitational waves emitted by intermediate- and extreme-mass-ratio inspirals (IMRIs/EMRIs), offering a promising way to indirectly detect dark matter. This detection channel, however, requires the spike to survive over cosmological time-scales. We investigate the impact that a host black hole's merger history has on its dark-matter spike. We focus on how the inspiral of a massive-black-hole (MBH) binary modifies and disrupts these spikes within the central $\sim 100$ Schwarzschild radii of the black hole. We find that a single merger with mass ratio $q = 0.03$ is capable of reducing a spike's density by around 3 (5) orders of magnitude at 100 (5) Schwarzschild radii across all polar angles relative to the binary's plane, independent of the primary MBH's mass, with larger mass ratios inducing even stronger depletion. In fact, a single past merger with mass ratio as small as $q = 0.005$ reduces central densities by 2--3 orders of magnitude, making gravitational-wave dephasing difficult to detect. We relate these results to the MBH merger history by analysing cosmological simulation data from {\tt Horizon-AGN}. We find that mergers become increasingly common as the MBH mass increases, with mergers of mass ratio $q \geq 0.03$ having occurred in $\sim 50$\% of MBHs of mass $3 \times 10^{7} \, \Msolar$ and $\sim 95\%$ of mass $10^{8} \, \Msolar$ at $z = 0$. These fractions are reached at even lighter masses ($2 \times 10^7 \, \Msolar$ and $5 \times 10^7 \, \Msolar$, respectively) at $z = 3$. We find that accretion-induced adiabatic MBH growth is unable to cause appreciable spike re-growth between such mergers in {\tt Horizon-AGN}. Thus, in combination with previous studies of EMRI-induced depletion, this work constrains the mass range over which massive black holes can host detectable DM spikes to $10^6 \, \Msolar$--$5 \times 10^7 \, \Msolar$ for $z \leq 3$.
\end{abstract}


\color{black}
\section{Introduction}
\label{sec:Introduction}
Dark matter (DM) is a fundamental part of the standard model of cosmology, but its exact nature remains one of the fundamental mysteries in modern physics. As part of the decades-long search for its properties, \citet[][henceforth \GS]{Gondolo_1999} have suggested that intermediate-mass and super-massive black holes (IMBHs/SMBHs) offer unique environments around which extremely dense dark-matter concentrations can form, providing opportunities to detect dark matter \citep{FORNASA_2008, Bertone_2020} and study it. Understanding how such environments evolve after formation may be crucial for understanding dark matter and its role in the evolution of the Universe.

\GS demonstrated that extremely high dark-matter concentrations, called dark-matter \textit{spikes}, can form around massive black holes (MBHs; a term which we use to refer to any black hole heavier than $10^4 \, \Msolar$) that are adiabatically grown within a pre-existing dark-matter cusp (e.g. the inner cusp of a dark-matter halo). In particular, if the pre-existing DM cusp had an initial density profile $\rho_{\rm DM} \propto r^{-\gamma_0}$, the spike's final profile is $\rho_{\rm DM} \propto r^{-\gamma_{\rm sp}}$ with
\begin{align}
    \gamma_{\rm sp} = \frac{9 -2 \gamma_0}{4 - \gamma_0}.
\end{align}
The density is usually normalised such that $\rho_{\rm DM}(r_{\rm sp}) = 226 \, \Msolar/\pc^3$, where $r_{\rm sp} = 0.2 r_{\rm SoI}$ \citep{Bertone_2005a}, with $r_{\rm SoI}$ denoting the radius of the sphere of influence around the MBH. Remarkably, $\gamma_0 \in (0,2)$ gives $\gamma_{\rm sp} \in (2.25, 2.5)$ -- a very narrow range; such a profile can produce DM densities up to $10^{12} \, \Msolar/\pc^3$ within the central few Schwarzschild radii of MBHs of mass $\mMBH = 10^{10} \Msolar$, and $10^{19} \, \Msolar/\pc^3$ for $\mMBH = 10^4 \, \Msolar$.

Some theories of self-interacting and self-annihilating DM predict that such large densities can produce detectable neutrino and gamma-ray fluxes, initially motivating extensive studies \citep{Ullio_2001, Gnedin_2004, Merritt_2002, Merritt_2004, Merritt_2007, Egorov_2026}. It was later shown that even collisionless DM spikes may be detected using gravitational waves (GWs) by inducing detectable GW dephasing in intermediate- and extreme-mass-ratio inspirals (IMRIs/EMRIs) \citep{Eda_2013, Sadeghian_2013,Eda_2015, Yue_2018, Yue_2019a, Yue_2019b, Hannuksela_2020, Kavanagh_2020, Cole_2022, Dai_2022, Becker_2022, Coogan_2022, Speeney_2022, Dosopoulou_2024, Montalvo_2024, Fischer_2024, Mukherjee_2024, Kavanagh_2024, Karydas_2024, Mitra_2025, Vicente_2025, Acevedo_2025, Karydas_2025, Speri_2026, Chen_2026, Wade_2026, Wu_2026}. Additionally, they may affect the inspiral of near-equal-mass binaries, leaving potentially observable imprints on the stochastic GW background \citep{Ghoshal_2023, Alonso_Alvarez_2024, Hu_2025, Chen_2025, Shen_2025b, Shen_2025a, Tiruvaskar_2026}, cause detectable changes to the flavour composition of neutrinos reaching Earth from active galactic nuclei \citep{Dev_2026}, and possibly explain galaxy rotation curve anomalies \citep{Herrera_2026a}. Recent work has also suggested using GW dephasing to measure the particle mass of ultralight scalar DM \citep{Vicente_2026}. 

While those predictions are promising and interesting, the spike's evolution on cosmological time-scales, which is often neglected, can have a significant impact on its detectability. \citet{Gnedin_2004} and \citet{Merritt_2004} demonstrated that, outside of $\sim 10^{-3}$--$10^{-2}$ pc from the central MBH, two-body relaxation of DM with a single-mass stellar cluster drives the DM density profile towards an $r^{-3/2}$ power-law profile, characteristic of a light species interacting with a more massive one in a Bahcall--Wolf profile \citep{Bahcall_1976, Bahcall_1977}. It was later shown that this relaxation is sped up significantly within multi-mass stellar cusps due to mass segregation \citep{Mukherjee_2024, Sharpe_2026a}. \citet{Herrera_2026b} have found that non-ideal MBH formation scenarios, such as an off-centre MBH seed or non-zero initial mass, can have a sizeable impact on the resulting DM over-density -- an idea first pointed out by \citet{Ullio_2001}. Further, \citet{Sharpe_2026a, Sharpe_2026b} have shown that over $\sim \Gyr$ time-scales, repeated EMRIs coming from the surrounding stellar cluster deplete spike densities by multiple orders of magnitude, plausibly making them undetectable by the Laser Interferometer Space Antenna (LISA) for MBHs with masses of roughly $\lesssim 10^6 \, \Msolar$, even for a modest EMRI rate of $\sim 100$--$300 \, \Gyr^{-1}$ per MBH. Heavier MBHs, however, are less susceptible to EMRIs due to the smaller mass ratio (for a fixed stellar-mass BH mass) preventing significant damage being done to the spike. This has been shown to be consistent with other studies \citep[e.g.,][]{Karydas_2026} which predict slightly different depletion fractions \citep[see][Appendix G]{Sharpe_2026a}. 

In this paper, we consider another local environmental effect on dark-matter spikes: mergers with other MBHs. Early studies focused on how near-equal-mass MBH binaries with separations on parsec scales deplete DM halos \citep{Milosavljevi_2001, Merritt_2002}; but one natural question that follows is how DM spikes respond to MBH mergers on \textit{sub}-parsec scales and below -- namely, the central $\sim 100$ Schwarzschild radii. In particular, is an individual major merger ($0.3 \leq q \leq 1$ with $q$ denoting the mass ratio), minor merger ($0.1 \leq q < 0.3)$, or even intermediate-mass-ratio merger ($10^{-3} \lesssim q < 0.1$) capable of depleting a DM spike significantly? Additionally, can accretion-driven adiabatic MBH re-growth between individual MBH mergers resurrect the spike, replenishing the depleted DM? These are questions we will endeavour to address below.

Throughout this paper, we will use LISA's one-radian detectability threshold \citep{Hinderer_2008} as a fiducial reference benchmark for detectability across all masses up to and beyond $10^9 \, \Msolar$, despite EMRIs orbiting such MBHs typically being outside of the LISA band. The true detectability threshold would depend on the waveform systematics, the signal-to-noise ratio, the observation time, and the detector itself, yet this serves as an approximate benchmark and allows for an order-of-magnitude estimate.

To understand the merger history of MBHs, we turn to cosmological simulations. Such simulations use first principles to provide self-consistent models of galaxy and MBH evolution. We will use {\tt Horizon-AGN} \citep{Dubois_2016}, a large-volume $(100 \ h^{-1} \, \rm Mpc)^3$ -- with $h$ the dimensionless Hubble constant -- cosmological simulation that tracks the dynamical evolution of MBHs and their resulting black-hole merger histories on the fly. This simulation has been shown to predict observed geometric properties of galaxies accurately \citep{Dubois_2014,Dubois_2016}, as well as the redshift, $z$, and luminosity evolution of the intrinsic alignments of galaxies \citep{Chisari_2015, Chisari_2016, Codis_2018}, the galaxy stellar mass function \citep{Beckmann_2017, Laigle_2017}, galaxy density profiles \citep{Peirani_2018}, and MBH population properties \citep{Volonteri_2016, Martin_2018}.

We therefore address two related questions. First, what binary mass ratio is sufficient for a \emph{single} MBH merger to reduce the central spike density such that any induced dephasing is below the one-radian benchmark?\footnote{We are setting an upper bound on this threshold mass ratio as MBHs can undergo multiple such mergers, reducing the densities even further.} Secondly, how frequently do MBHs of different masses experience such mergers? We answer the first using direct three-body simulations and the second using {\tt Horizon-AGN} merger histories. This, in combination with the previous work of \cite{Sharpe_2026b}, will further restrict the viable mass range over which MBHs could potentially host spikes that are detectable via GW dephasing of IMRIs/EMRIs. Common parameters used throughout the text are summarised in Table \ref{tab:Parameters}.

The rest of this paper is organised as follows. In \S\ref{sec:Disruption of a Spikes by Individual Mergers}, we model the depletion of a collisionless DM spike induced by a single MBH merger as a function of the merger's mass ratio. In \S\ref{sec:Detectability of DM Spikes}, we use simple dephasing calculations to examine the detectability prospects of the DM that remains after a merger. In \S\ref{sec:MBH Merger Histories}, we study MBH merger trees to understand merger histories as a function of mass ratio, and hence the extent to which one should expect DM spikes to have been depleted by MBH mergers. We then synthesise and discuss our results in \S\ref{sec:Discussion}, before concluding in \S\ref{sec:Conclusion}.

\section{Disruption of Dark-Matter Spikes by Individual Mergers}
\label{sec:Disruption of a Spikes by Individual Mergers}

To determine the depletion of collisionless DM induced by a single MBH merger, we integrate the evolution of DM test particles in the presence of an inspiralling MBH binary system. We assume an initially thermal distribution of DM eccentricities, with a distribution function $f_{\eDM} = 2\eDM$ with $\eDM$ denoting the DM particle's eccentricity, and an isotropic density profile given by 
\begin{align}
    \rho_{\rm DM} = \rho_{\rm sp} \left(\frac{r}{r_{\rm sp}}\right)^{-\gamma_{\rm sp}}\,, \label{eq:rho_DM}
\end{align}
as mentioned in the introduction,
where $\rho_{\rm sp} = 226 \,\Msolar/\pc^3$ and $r_{\rm sp} = 0.2 \,r_{\rm SoI}$ with
\begin{align}
    r_{\rm SoI} = 1.474 \left(\frac{\mMBH}{10^6 \, \Msolar} \right)^{0.543} \ {\rm pc}\,. \label{eq:r_SoI}
\end{align}
This scaling of $r_{\rm SoI}$ with $\mMBH$ comes from the definition $r_{\rm SoI} \equiv \mathrm{G}\mMBH/\sigma^2$ where the velocity dispersion $\sigma$ is given by the $M$--$\sigma$ in \citet[Eq.~7]{Kormendy_2013}. We select the commonly studied $\gamma_{\rm sp} = 7/3$, as predicted by \GS for an initially Navarro--Frenk--White (NFW; \citealt{NFW1997}) DM cusp. (Note that $\gamma_{\rm sp}$ can range from $2.25$--$2.5$ and, as such, this choice is simply a fiducial value.) 

We wish to obtain two quantities: the fraction of DM particles that survive the inspiral without being ejected or swallowed by one of the MBHs, and the distribution of these left-over DM particles. This will allow us to construct post-merger density profiles, which then determine the DM-induced observational signal. We take the masses of the primary and secondary MBHs to be $\Mp$ and $\Ms = q \Mp \leq \Mp$, respectively, where $q$ is the mass ratio, with total mass $\Mps = \Mp + \Ms$. We ignore reductions in rest mass due to GW energy losses,\footnote{These are, at most, about 5\% for non-spinning MBHs and 11.4\% for maximally spinning MBHs, corresponding to a $q = 1$ merger \citep{Jim_nez_Forteza_2017}.} and take their Schwarzschild radii to be $\Rschp$ and $\Rschs$, respectively, with the radius of the remnant MBH being $\Rschps$. We now develop our model for studying this system.

\begin{table}
    \centering
        \def\sep{0.1ex}
        \begin{tabular}[c]{|c|c|}
            \hline 
            Symbol & Definition \\ 
            \hline
            $\Mp$ & Mass of the primary MBH \\ [\sep]
            $\Ms$ & Mass of the secondary MBH \\ [\sep]
            $q$ & $\Ms/\Mp$ \\ [\sep]
            $\Mps$ & $\Mp + \Ms$ \\ [\sep]
            $\ME$ & Mass of the subsequent secondary EMRI component \\ [\sep]
            $\mMBH$ & Generic MBH mass \\ [\sep]
            $\Rschp$ & Schwarzschild radius of the primary MBH \\ [\sep]
            $\Rschs$ & Schwarzschild radius of the secondary MBH \\ [\sep]
            $\Rschps$ & $\Rschp + \Rschs$ \\ [\sep]
            $\rpDM$ & DM particle's inner turning point ($\equiv$ periapsis) \\ [\sep]
            $\raDM$ & DM particle's outer turning point ($\equiv$ apoapsis) \\ [\sep]
            $\eDM$ & DM particle's eccentricity ($\equiv (\raDM - \rpDM)/(\raDM + \rpDM)$) \\ [\sep]
            $\aDM$ & DM particle's semi-major axis ($\equiv \rpDM/(1-\eDM)$) \\ [\sep]
            $\aMBHB$ & Semi-major axis of the MBH binary \\ [\sep]
            $\aEMRI$ & Semi-major axis of the EMRI \\ [\sep]
            $\iota$ & Relative inclination between DM orbit and MBH binary \\ [\sep]
            \hline
        \end{tabular}
        \caption{Summary of common parameters used in this work.}
    \label{tab:Parameters}
\end{table}

\subsection{Our Three-body Model}
\label{subsec:Our Three-body Model}

We assume that the binary has an initial eccentricity of zero -- the true distribution of MBH binary eccentricities is uncertain as it strongly depends on the local environment as well as the inspiral stage \citep{Sesana_2010, Zrake_2021, Gualandris_2022, Siwek_2024, Tomaselli_2026uqg}. Further, it has been shown that MBH spin only significantly impacts the DM spike within the central $\sim 5$ Schwarzschild radii \citep{Mitra_2025}. We do not consider distances smaller than this and thus take each MBH to be non-spinning. We then run the binary all the way down to merger using 2.5 post-Newtonian (PN) orbital corrections and consider the binary merged when the components' horizons overlap. Note that this is unlike the semi-analytic approach for modelling EMRI-induced DM depletion presented in \citet{Sharpe_2026b}, where we modelled a single orbit and took the probability of ejection of a DM particle during a single binary orbit to be independent of the binary's previous orbits. Here, the entire evolution is computed numerically.  
Due to the DM being collisionless and the total DM enclosed within the central $\mathcal{O}(100) \Rschps$ (the distances considered here) being much smaller than either $\Mp$ or $\Ms$, we assume the DM particles behave independently of one another; that is, they only experience the gravitational influence of the binary components but not of each other. Thus, we study the response of the DM spike by separately modelling the response of \textit{individual} DM particles. In this paper, we run $\sim 3 \times 10^7$ Monte Carlo three-body DM--MBH--MBH full-inspiral simulations to obtain a statistical understanding of the depletion.

Due to the rotational invariance of our system, it is only the relative inclination between the binary and the DM particle that determines the evolution, as opposed to both inclinations separately. Hence, for each simulation, we set the binary's initial inclination to zero, randomise the DM particle's initial inclination, $\iota_0$, uniformly in $\cos \iota_0$, and randomise the mean anomaly. We also randomise the DM particle's initial longitude of the ascending node, $\Omega_0$, and initial longitude of periapsis, $\varpi_0$, between $0$ and $2\pi$.

Interestingly, the evolution of an MBH-binary inspiral does not depend on each mass separately but only on its mass ratio $q$, so long as distance and time are measured in units of $\Rschps$ and $\Rschps/c$, respectively (this is proven in Appendix \ref{appsec:Over-all Mass Independence of Orbital Evolution}). Hence, we will treat only $q$ as a free parameter, while fixing $\Mp = 10^6 \, \Msolar$ -- this choice is completely arbitrary and any other selection would leave the results of this section unchanged. One can then generalise to any $\Mp$ and $\Ms$ by simply adjusting the mass ratio.

Finally, there are three possible states the DM particle can be in at the end of a run: bound, unbound, or swallowed by one of the MBHs. We consider a particle swallowed once it passes within one Schwarzschild radius of an MBH's centre. Included in our definition of swallowed are any DM particles whose bound orbit puts it on a direct plunge trajectory into the remnant MBH. Assuming the final binary-merger product is a single MBH, we classify a DM particle as bound (unbound) if its relativistic specific energy at the time of merger obeys $E < \cl^2$ ($E \geq \cl^2$). Due to the remnant MBH absorbing the orbital angular momentum of the binary, it will possess a non-zero spin. However, in Appendix \ref{appsec:Escape velocities of (non-)rotating MBHs}, we show that even a maximally spinning MBH produces a negligible change in the required escape velocity of DM particles and hence in the final post-merger DM distribution. For simplicity, we will thus take the remnant to be non-spinning. A DM particle's relativistic specific energy is thus given by \citep{Misner_1973}
\begin{align}
    E^2 = \frac{\alpha^3 \mathrm{c}^6}{-v_r^2 + \alpha(\alpha \mathrm{c}^2 - v_\perp^2)}\,,
\end{align}
where $\alpha = 1 - \Rschps/r$ and $v_r$ and $v_{\perp}$ are the Schwarzschild coordinate radial and transverse velocities as measured by an observer at infinity.

\subsection{Monte Carlo Simulations}
\label{subsec:Our Monte Carlo Simulations}

We integrate our system using direct $N$-body simulations with {\tt multistar} \citep{multistar_ref} -- an $N$-body code containing 2.5PN piecewise corrections. We wish to obtain the probability of a DM particle surviving the inspiral as a function of its initial periapsis $\rpDMz$, initial eccentricity $\eDMz$, and initial inclination, $\iota_0$, as well as the distribution $f_{\rpDM, \eDM, \iota}$ of the final periapsis $\rpDM$, final eccentricity $\eDM$, and final inclination $\iota$ of surviving particles. 

However, initialising the orbits to have the desired eccentricities and periapsides requires care. Relativistic orbits are no longer Keplerian and hence there is no unique definition of the 6 standard orbital elements. A natural alternative is to assume the DM particle is following a Schwarzschild geodesic\footnote{Note that \multistar only contains 2.5PN corrections and hence particles will not follow exact Schwarzschild geodesics. However, this treatment serves as a very good approximation.}, specify the desired initial inner and outer turning points, interpret these as the initial periapsis $\rpDMz$ and apoapsis $\raDMz$, and calculate the initial eccentricity accordingly through $\eDMz = (\raDMz - \rpDMz)/(\raDMz + \rpDMz)$. All the relevant details can be found in \citet[][Appendix C]{Sharpe_2026a}, including the conversion between harmonic coordinates -- those used by {\tt multistar} -- and Schwarzschild coordinates. 

We discretise our $(\rpDMz, \eDMz)$ parameter space to obtain an expression for $f_{\rpDM, \eDM, \iota}$  for each $q$ numerically. We select $q = \{$0.005, 0.03, 0.1, 0.2, 0.4, 0.75$\}$, $\rpDMz/\Rschps = \{$5, 10, 20, 40, 60, 100$\}$, and $\eDMz = \{$0.1, 0.5, 0.9$\}$. We assign particles with $\rpDM \leq 5 \, \Rschps$ to the $\rpDM = 5 \, \Rschps$ cell, and particles with eccentricity below 0.1 or above 0.9 to the corresponding edge bins.

For the MBH binary, the assumption of circularity allows us to directly calculate the 2PN-corrected initial angular velocity \citep[Eq. (190)]{Blanchet_2002}
\begin{align}
    n_{\rm 2PN} = \sqrt{\frac{\G \Mps}{\aMBHBz^3} \left[1 + (\eta - 3) \epsilon + \left(6 + \frac{41}{4} \eta + \eta^2\right) \epsilon^2\right]}\,, \label{eq:n_2PN}
\end{align}
where $\eta = \Mp \Ms/\Mps^2$, $\epsilon = \G\Mps/(\cl^2 \aMBHBz)$ and $\aMBHBz$ is the initial MBH binary semi-major axis.

The domain of mass ratios and initial MBH binary semi-major axes we consider is limited by the computational cost of each three-body simulation. The cost scales with the number of orbits completed during the GW-induced inspiral, given by
\begin{align}
    N_{\rm orb}(\aMBHBz) &= \int_{\aMBHBz}^{0} \frac{\left<n_{\rm 2PN}\right>}{2\pi} \frac{1}{\langle\dot \aMBHB \rangle} \dd \aMBHB \\
    &\approx \frac{\aMBHBz^{5/2} \cl^5}{64\pi \G^{5/2} \Mp^{5/2} q \sqrt{1 + q}}\left(1 + \mathcal{O}(\epsilon) + \mathcal{O}(\epsilon^2) + \hdots\right) \,, \label{eq:N_orbs}
\end{align}
where we have used \citep{Peters_1964}
\begin{align}
    \langle \dot{\aMBHB} \rangle &= - \frac{64}{5} \frac{\G^3 \Mp \Ms \Mps}{\cl^5 \aMBHB^3}\,.\label{eq:adot GW}
\end{align}
For computational efficiency, we limit the initial separation distance to $\aMBHBz \leq 100\,\Rschps$ and only consider mass ratios down to $q = 0.005$. Further, when the DM particle's periapsis and apoapsis are much smaller than $100\,\Rschps$, we initialise the MBH binary with a smaller semi-major axis that still obeys $\raDMz < \aMBHBz \leq 100\,\Rschps$ where $\raDMz$ is the DM particle's initial apoapsis. For the purposes of reproducibility, we state the exact choices in Appendix~\ref{appsec:Initial MBH Binary Semi-major Axes}.

Finally, the DM particle is initialised in orbit around the primary and therefore inherits the primary's barycentric motion. Since the primary receives an initial kick as it is forced into orbit with the secondary, the DM particle receives the same initial velocity kick. At large DM-primary separations, however, this breaks down and can lead to the DM particle receiving a kick much larger than its orbital velocity. Thus, for initial DM-primary separations greater than $150\,\Rschps$, the DM particle's orbit is instead referenced to the binary's centre of mass, preventing this unphysical kick. While the choice of $150\,\Rschps$ is arbitrary, we have numerically confirmed our results are insensitive to its exact value.

On the short length-scales considered here, any ejection of a dark-matter particle, induced by the binary, is irreversible. The DM cannot relax with stars -- due to the total enclosed stellar mass falling below that of a single star, leaving few to no stars to relax with -- or through self-relaxation due to the extremely small individual particle mass, which makes the two-body relaxation time-scale $\sim 10^{70}$ yr \citep{Kavanagh_2020}. On parsec scales, however, such as those considered by \citet{Ullio_2001, Milosavljevi_2001, Merritt_2002, Merritt_2007,Sharpe_2026a,Sharpe_2026b}, there are sufficient stars for the depleted phase space to be at least partially refilled through the formation of a Bahcall--Wolf cusp -- the corresponding DM densities, however, are much lower than that of the original spike and continue to decrease due to continual heating.

\subsection{Calculating the Post-Merger Density Profile}
\label{subsec:Calculating the Post-Merger Density Profile}

Due to the azimuthal symmetry of our system, we expect the final density profile only to depend on the radius $r$ and the polar angle $\phi$ measured from the positive $z$-axis. We verified numerically that indeed the remaining distribution is axisymmetric.

Since we are working with a finite number of DM particles, the post-merger DM distribution is given by a weighted sum over all simulated particles that remain bound, \emph{viz.}
\begin{align}
    \begin{aligned}
        f_{\rpDM, \eDM, \iota} = &\sum_i^{N_{\rm bound}(q)} w(q, \rpDMz^{(i)}, \eDMz^{(i)})  \\
        & \ \ \ \times \delta((\rpDM, \eDM, \cos \iota) - ({\rpDM}^{(i)}, {\eDM}^{(i)}, \cos {\iota}^{(i)}))\,,
    \end{aligned} \label{eq:f_rp,e,i}
\end{align}
where $\delta$ is the Dirac delta function, $N_{\rm bound}(q)$ is the number of remaining bound DM particles, and $\rpDMz^{(i)}, \eDMz^{(i)}, {\rpDM}^{(i)}, {\eDM}^{(i)}$ and ${\iota}^{(i)}$ are particle $i$'s initial periapsis, initial eccentricity, final periapsis, final eccentricity, and final inclination, respectively. The weights, $w$, account for the fact that the initial density of DM particles in $(\rpDMz, \eDMz)$ space is not uniform and that $N_{\rm bound}$ will depend on $q$ as well as the total number of simulated particles, $N_{\rm total}$. They are also independent of inclination as we already sample our DM particles uniformly in $\cos \iotaz$. These weights are given by
\begin{align}
    \begin{aligned}
        w(q, \rpDMz^{(i)}, \eDMz^{(i)}) &= \frac{A_w}{N_{\rm total}} \int_{\rpDM^-(i)}^{\rpDM^+(i)} \dd \rpDM \\
        &\times \int_{\eDM^-(i)}^{\eDM^+(i)} \dd \eDM \, 2 \eDM \rpDM^{2 - \gamma_{\rm sp}} (1 - \eDM)^{\gamma_{\rm sp} - 3},
    \end{aligned} \label{eq:weights}
\end{align}
with $A_w$ chosen such that the weights of all simulated particles sum to unity. Here, $\rpDM^{\pm}(i)$ represent the lower and upper midpoints between our discrete choices of variables for the periapsis bin within which particle $i$ is contained, with $\eDM^{\pm}(i)$ defined analogously for eccentricity. We have taken the initial spike density and eccentricity distribution to be $\rho_{\rm DM} \propto r^{-\gamma_{\rm sp}}$ and $f_{\eDM} = 2\eDM$, respectively.

We now split our $(r, \phi)$ space into cells, and calculate each bound particle's average contribution to that cell. This averaging procedure helps to reduce noise in our final density calculation. We denote the $j$th radial and $k$th polar-angle left cell boundaries by $r_j$ and $\phi_k$, respectively. The cell-averaged probability density of finding DM particle $i$, whose orbital parameters are $(\rpDM^{(i)}, \eDM^{(i)}, \iota^{(i)})$, inside cell $(j,k)$, i.e., with radius $r \in [r_j, r_{j+1})$ and polar angle $\phi \in [\phi_k, \phi_{k+1})$, is
\begin{align}
    K_{j, k}^{(i)} = \frac{1}{V_{j,k}}\int_{V_{j,k}} \dd V \,\varrho_{\rm DM}(r, \phi, \rpDM^{(i)}, \eDM^{(i)}, \iota^{(i)})\,,
\end{align}
where
\begin{align}
    \dd V = 2\pi \frac{r^2 \, \sin \phi}{{\sqrt{1 - \Rschps/r}}} \, \dd r \, \dd \phi
\end{align}
is the Schwarzschild proper volume element, $\varrho_{\rm DM}(r, \phi, \rpDM^{(i)}, \eDM^{(i)}, \iota^{(i)})$ is the probability density for DM particle $i$ being a distance $r$ from the MBH with polar angle $\phi$, and
\begin{align}
    V_{j,k} = 2\pi \left(\cos \phi_{k} - \cos \phi_{k+1}\right) \int_{r_{j}}^{r_{j+1}} \frac{r^2}{\sqrt{1 - \Rschps/r}} \dd r
\end{align}
is the volume of cell $(j,k)$. We then sum the contribution from each bound DM particle to obtain 
the density function of the fraction of all simulated DM particles that reside at $r, \phi$
\begin{align}
    \rho_{\rm rem}(q, r, \phi) &= \sum_i^{N_{\rm bound}} w(q, \rpDMz^{(i)}, \eDMz^{(i)}) \, K_{j, k}^{(i)} \,,
\end{align}
where $(j,k)$ represents the indices of the cell containing $(r, \phi)$. We choose 70 logarithmically spaced bins between $5\,\Rschps$ and $100\,\Rschps$ for $r$, and 60 equally spaced bins between $0$ and $\pi$ for $\phi$.

To calculate $\varrho_{\rm DM}$, consider a single DM particle at $(\rpDM, \eDM, \iota)$. We wish to find the probability that, at any given instant, it sits at a separation distance between $r$ and $r + \dd r$ and has a polar angle between $\phi$ and $\phi + \dd \phi$. By averaging over the azimuth (due to relativistic apsidal advance) and noticing that orbits in the Schwarzschild metric are planar, the two criteria become independent and can hence be considered separately. To calculate the former, we treat our azimuthally averaged orbit as a disc and simply use the probability density, $\sigma_r$, of finding a particle on an orbit with periapsis $\rpDM$ and eccentricity $\eDM$ at a radius $r$. This is given by
\begin{align}
    \sigma_r &= \frac{\left|\dd t/\dd r\right|}{\int_{\rpDM}^{\raDM} \left|\dd t/\dd r\right| \dd r}\,, \label{eq:sigma_r orbit}
\end{align}
where $\raDM$ is the particle's apoapsis and \citep{Sharpe_2026a}
\begin{align}
    \left|\frac{\dd r}{\dd t}\right| &= \frac{\cl^2(1 - \Rschps/r)}{E}\sqrt{\frac{E^2}{\cl^2} - \cl^2\left(1 - \frac{\Rschps}{r}\right)\left(1 + \frac{L^2}{\cl^2r^2}\right)}\,,\\
    E^2 &= \frac{\mathrm{c}^4(\rpDM - \Rschps)(\raDM - \Rschps)(\rpDM + \raDM)}{\rpDM \raDM (\rpDM + \raDM) - \Rschps(\rpDM^2 + \rpDM \raDM + \raDM^2)}\,,\\
    L^2 &= \frac{\mathrm{c}^2 \Rschps \rpDM \raDM}{(\rpDM + \raDM) - \Rschps(\rpDM/\raDM + 1 + \raDM/\rpDM)}\,.
\end{align}
The probability of the DM particle being found between $r$ and $r + \dd r$ is then $\sigma_r \, \dd r$. 

To calculate the polar-angle distribution, let us suppose our DM particle is orbiting in the $x$--$y$ plane, and at some instant is located at $(r \cos \omega, r \sin \omega, 0)$ for some angle $\omega \in [0, 2\pi)$. By rotating this orbit by an angle $\iota$ about the $x$-axis (we can rotate about any axis in the $x$--$y$ plane since the post-merger distribution is axially symmetric) such that its new position vector becomes $(r\cos \omega, r\cos \iota \sin \omega, r\sin \omega \sin \iota)$, we find $\cos \phi = \sin \omega \sin \iota$. Assuming apsidal precession randomises the longitude of periapsis of the orbit, we can treat $\omega$ as uniformly distributed in $[0, 2\pi)$. By then applying the change-of-variables formula for probability densities, we find that, for a given inclination, the distribution of polar angles is
\begin{align}
    f_\phi (\phi) &= 
    \begin{cases}
        \dfrac{1}{\pi} \dfrac{\sin \phi}{\sqrt{\sin^2 \iota - \cos^2 \phi}}, &\text{ if } |\cos{\phi}| < \sin \iota, \\
        0, &\text{ otherwise.}
    \end{cases} \label{eq:f_phi}
\end{align}
The probability of the DM particle being found with polar angle between $\phi$ and $\phi+ \dd \phi$ is then $f_\phi \, \dd \phi$. 

Thus the probability density of a single particle with orbital parameters $(\rpDM, \eDM, \iota)$ to reside at $(r, \phi)$ is 
\begin{align}
    \varrho_{\rm DM} &= \frac{\sigma_r \, \dd r \, f_\phi \, \dd \phi}{\dd V} = \frac{\sigma_r \, f_\phi \, \sqrt{1 - \Rschps/r}}{2\pi r^2 \sin \phi}\,, \label{eq:varho_DM}
\end{align}
if $|\cos{\phi}| < \sin \iota$ and $\rpDM < r < \raDM$, and zero otherwise.

Finally, to obtain the local fraction of the original density remaining, we normalise $\rho_{\rm rem}$ by dividing by the original pre-merger fractional density to get
\begin{align}
    \mathcal{F}_{\rm rem} (r, \phi) = \frac{\rho_{\rm rem}(q, r, \phi)}{\sum_i^{N_{\rm total}} w(q, \rpDMz^{(i)}, \eDMz^{(i)}) \frac{1}{V_{j,k}}\int_{V_{j,k}} \dd V \,\varrho_{\rm DM}^{(i,\,{\rm init)}}} \,, \label{eq:F_rem}
\end{align}
where $\varrho_{\rm DM}^{(i,\,{\rm init)}} = \varrho_{\rm DM}(r, \phi, \rpDMz^{(i)}, \eDMz^{(i)}, \iotaz^{(i)})$ and, again, $(j,k)$ represents the indices of the cell that $(r, \phi)$ resides in.

At each $(\rpDMz, \eDMz)$ grid point, we run $2.5 \times 10^4$ simulations for $q = 0.005$, $5.0 \times 10^4$ simulations for $q = 0.03$, $10^5$ simulations for $q = 0.1$, and $5 \times 10^5$ simulations for $q = \{$0.2, 0.4, 0.75$\}$. This gives the total number of three-body simulations per mass ratio, $q$, to be
\begin{align}
    N_{\rm total}(q) = 
    \begin{cases}
        4.5 \times 10^5, &\text{if } q  = 0.005, \\
        9.0 \times 10^5, &\text{if } q  = 0.03, \\
        1.8  \times 10^6, &\text{if } q = 0.1,\\
        9.0  \times 10^6, &\text{if } q = \{0.2, 0.4, 0.75\}.
    \end{cases} \label{eq:N_total(q)}
\end{align}
These larger mass ratios induce greater DM density reductions and thus require more particles to control finite-sampling noise. These values were chosen such that the maximum 1-$\sigma$ uncertainty in $\log_{10} \mathcal{F}_{\rm rem}$ at $\phi = \pi/2$ is less than $10\%$ -- these uncertainties were calculated using bootstrapping, as we describe in the next subsection. Away from $\phi = \pi/2$, however, uncertainties increase due to the condition $\sin^2 \iota > \cos^2 \phi$, in Eq.~\eqref{eq:varho_DM}, reducing the number of contributing particles, with the noise diverging as $\phi \to 0$. Nonetheless, the noise in the lowest bin is finite and remains controlled.

\subsection{Numerical Results}
\label{subsec:Numerical Results}

\begin{figure}
    \centering
    \includegraphics[width=0.49\textwidth]{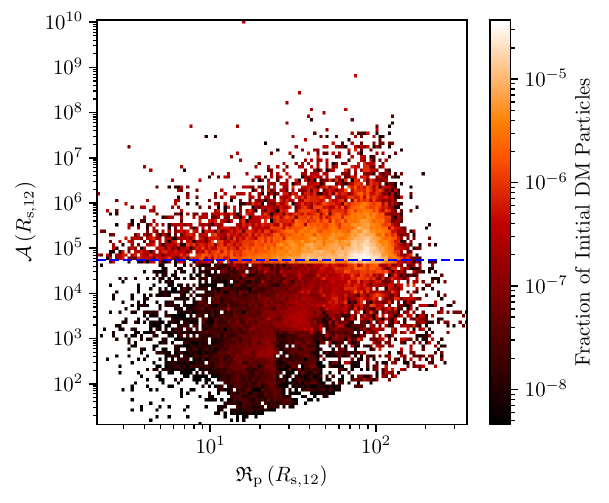}
     \caption{
     The distribution, weighted by $w$ (Eq.~\ref{eq:weights}), of periapsides $\rpDM$ and semi-major axes $\aDM$ of DM particles that remain bound after a merger of mass ratio $q = 0.2$. Most remaining DM particles sit on extremely wide, weakly bound orbits. The blue horizontal dashed line represents the critical semi-major axis predicted by Eq.~\eqref{eqn:A_crit}.
     }
    \label{fig:rp_a_2D_hist_0_2}
\end{figure}

Figure \ref{fig:rp_a_2D_hist_0_2} shows the distribution of the surviving DM particles as a function of the periapsis, $\rpDM$, and semi-major axis, $\aDM$, after a merger of mass ratio $q = 0.2$ weighted by $w$ (Eq.~\ref{eq:weights}). We note that the large relative contributions attributed to individual isolated cells are due to these weights spanning a wide range of magnitudes.

\begin{figure*}
    \centering
    \includegraphics[width=0.99\textwidth]{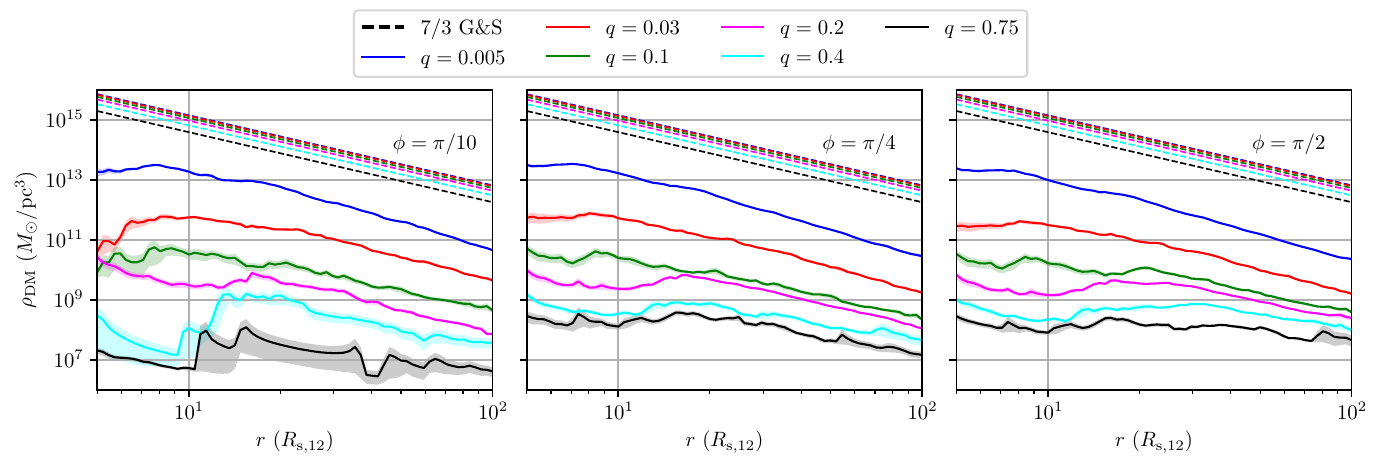}
    \caption{
    The DM density profile around a remnant MBH whose primary had a mass of $\Mp = 10^6 \, \Msolar$ after a merger with mass ratio $q = \{0.005, 0.03, 0.1, 0.2, 0.4, 0.75\}$ for $\phi = \pi/10$ (left), $\phi = \pi/4$ (centre), and $\phi = \pi/2$ (right). The shaded regions represent the 68\% (1-$\sigma$) confidence bands. For each mass ratio, we also show the original \GS profile (dashed). Recall the \textit{fractional} reduction in density is independent of the MBH mass if distance is measured in units of the Schwarzschild radius.
    }
    \label{fig:Post_merge_profiles_phis_cell_mean}
\end{figure*}

There is a clear horizontal cliff at $\aDM \approx 5.4 \times 10^4 \, \Rschps$ for $\rpDM \lesssim 100 \, \Rschps$. This arises because we initialise our binary no further out than $100 \, \Rschps$. If a DM particle undergoes a strong interaction that places it on an orbit whose period is longer than the GW inspiral time, given by \citep{Peters_1964}
\begin{align}
    T_{\rm GW} = \frac{5 \cl^5}{256 \G^3} \frac{(100 \, \Rschps)^4}{\Mps \Mp \Ms},
\end{align}
then the DM particle will not undergo any further strong interaction as the binary will have merged by the time it reaches periapsis again. Such a particle is hence safe and remains bound at the end of the run. This introduces a critical semi-major axis, given by
\begin{equation}\label{eqn:A_crit}
    \aDM_{\rm crit} = \left(\frac{G\Mps T_{\rm GW}^2}{4\pi^2}\right)^{1/3}\,,
\end{equation}
above which DM particles are guaranteed to not be ejected by a strong three-body interaction. For a binary of mass ratio $q = 0.2$ that is separated by $100\,\Rschps$, the corresponding critical semi-major axis is $\aDM_{\rm crit} = 5.432 \times 10^4 \, \Rschps$, exactly as seen in Figure \ref{fig:rp_a_2D_hist_0_2}.

This critical semi-major axis would be pushed to larger distances if we initialised our MBH binary further out, suggesting that these particles falsely contribute to the DM densities. However, they provide a negligible contribution (typically $\sim 0.1\%$--$1\%$) within the central $100\,\Rschps$ -- the region we are studying. This is because they spend almost all of their orbital period near apoapsis with their contribution to the central density scaling as $\sim (1 - \eDM)/\sqrt{\raDM} \propto \aDM^{-3/2}$ when $\aDM \gg \rpDM$ (Eq.~\ref{eq:sigma_r orbit}). Thus, increasing the binary's initial separation distance could only either strengthen the depletion or leave results effectively unchanged within the radii considered, while significantly increasing computational cost (see \S\ref{subsec:Our Monte Carlo Simulations}). We hence keep our initialisation at $100\,\Rschps$.

Figure \ref{fig:Post_merge_profiles_phis_cell_mean} shows the radial density profile, calculated using Eqs. \eqref{eq:rho_DM} and \eqref{eq:F_rem}, after a merger whose primary has a mass of $\Mp = 10^6 \, \Msolar$ as a function of the mass ratio and polar angle, $\phi$ along with 68\% (1-$\sigma$) confidence bands calculated by dividing all runs into 20 random blocks and performing $10^3$ bootstrapping repetitions.\footnote{Bootstrapping calculates confidence intervals by repeatedly resampling data with replacement and calculating the mean. This is performed a large number of times and the 68\% confidence interval is bounded by the 16th and 84th percentiles of those averages.} The \GS profiles vary with $q$ since the remnant mass, given by $\Mps = \Mp(1+q)$, increases with $q$ and hence so does $\Rschps$. We see significant depletion all the way down to the smallest mass ratio considered, $q = 0.005$. We emphasise that the fractional reductions in density depend only on the mass ratio, rather than the individual masses. Substituting Eq.~\eqref{eq:r_SoI} into Eq.~\eqref{eq:rho_DM}, we find that the overall density normalisation itself scales as $\rho_{\rm DM} \propto \Mp^{-1.066}$ at a fixed distance, measured in Schwarzschild units, from the MBH.

We see a clear separation of ejection fractions over $q \in [0.005,0.75]$. However, when depletion fractions are large, only a small number of particles contribute meaningfully to the central densities, leaving the densities vulnerable to shot noise. This is particularly evident for high mass ratios and low polar angles, e.g., $q = 0.75$ and $\phi = \pi/10$, where the uncertainties are large. Accurately resolving such large density reductions simply requires simulating more particles.

It is natural to ask how additional mergers would further change the DM spike density profile. However, modelling this is difficult using direct $N$-body simulations. The maximum distance at which we could initialise our MBH binary was bottlenecked by computational cost. This meant that we could only reliably model the evolution of DM particles whose semi-major axis was within, or close to, $100\,\Rschps$. Particles further out will thus have their ejection probabilities underestimated. After the merger, almost all DM particles have enormous semi-major axes, as seen in Figure \ref{fig:rp_a_2D_hist_0_2}, and since we cannot initialise our binary further out than $100\,\Rschps$, we cannot model how a second merger would modify the profile accurately. However, it is likely only to deplete the spike even further.

We now turn to the detectability of these post-merger DM spikes, asking whether their reduced density can still produce a significant dephasing in future EMRI observations. 

\section{Detectability of DM Spikes After an MBH Merger from GW Dephasing}
\label{sec:Detectability of DM Spikes}

GW dephasing is one of the primary observational signatures of DM spikes \citep{Bertone_2020}. A spike surrounding an MBH modifies the orbital evolution of the inspiralling compact object, causing it to depart from the vacuum prediction of general relativity, by introducing measurable phase shifts in the emitted GW signal. EMRIs are particularly promising as they inspiral very slowly, allowing small effects to accumulate into a detectable phase shift. Here, we will provide a crude estimate of the DM-induced GW dephasing to highlight the impact of MBH mergers on DM detection prospects.

\begin{figure*}
    \centering
    \includegraphics[width=0.99\textwidth]{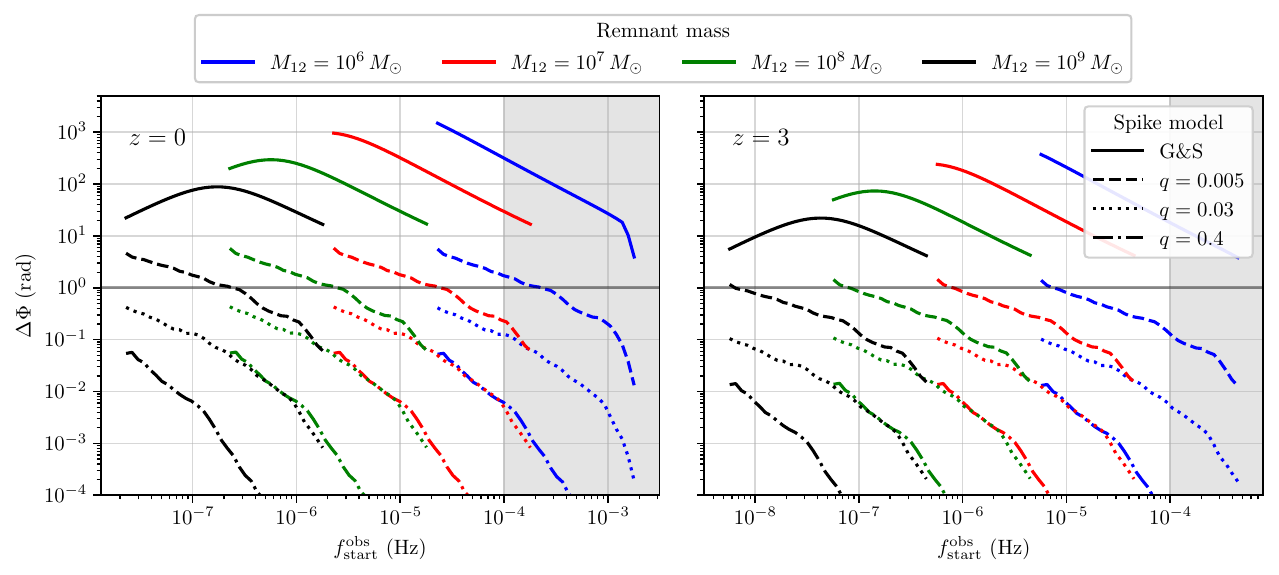}
    \caption{
    The dynamical-friction-induced GW dephasing for an EMRI with secondary mass $10 \, \Msolar$ in the plane of the original MBH-MBH merger ($\phi = \pi/2$) and varying remnant masses $10^6 \, \Msolar$ (blue), $10^7 \, \Msolar$ (red), $10^8 \, \Msolar$ (green), and $10^9 \, \Msolar$ (black) after the primary has undergone a merger of mass ratio $q = \{0.005, 0.03, 0.4\}$ for redshifts $z = 0$ (left) and $z = 3$ (right). This is compared to the undepleted \GS case (solid line). The initial observed GW frequency, $f_{\rm start}^{\rm obs}$, is chosen such that the initial separation distances range from $5\,\Rschps$ to $100\,\Rschps$. The LISA band is shaded grey for reference.
    }
    \label{fig:Dephasing_z_mult_cell_mean}
\end{figure*}
EMRIs emit predominantly in the mHz frequency band, making them promising targets for next-generation space-based GW observatories such as LISA, which is sensitive to GWs in the frequency range $10^{-4}$--$10^{-1}$ Hz \citep{Colpi_2024}, as well as other future low-frequency GW detectors. As stated previously, we take LISA's one-radian dephasing detectability threshold as a general benchmark for significant departure from the vacuum-GR waveform \citep{Hinderer_2008}. We focus on MBHs with masses $\geq 10^6\,\Msolar$, as these are among the systems most likely to have undergone mergers with other MBHs, as we show in \S\ref{sec:MBH Merger Histories}, and have not been disfavoured by EMRI-induced DM depletion \citep{Sharpe_2026b}.

The dominant form of DM-induced energy dissipation -- and hence dephasing -- of EMRIs is dynamical friction, particularly at the separation distances considered here \citep{Karydas_2024}. Now, the energy dissipation due to dynamical friction scales linearly with the local DM density \citep{Chandrasekhar_1943,Binney_2008,Mitra_2025}. However, we saw in \S\ref{sec:Disruption of a Spikes by Individual Mergers} that the DM density profile is not isotropic after a merger, particularly for $q \geq 0.2$ where the densities are greatest at $\phi = \pi/2$, and decrease for smaller $\phi$. We will assume all EMRIs inspiral within the $\phi = \pi/2$ plane. This maximises the DM-induced dephasing and hence provides a conservative estimate of the extent to which mergers reduce a spike's detectability.
Let us suppose an EMRI's primary component is a post-merger MBH with mass $\Mps$, while the secondary is a stellar-mass BH whose mass we take to be $\ME = 10 \, \Msolar$, as typically predicted by EMRI studies \citep[see, e.g.,][and references therein]{Babak_2017}. Assuming the EMRI remains quasi-circular during the inspiral, the total observed GW phase accumulated as its GW frequency $f$ in the source frame -- which is twice the orbital frequency -- evolves from $f^{\rm src}_{\rm start}$ to $f^{\rm src}_{\rm end}$ is
\begin{align}
    \Phi_{\rm total} = 2 \pi \int_{f^{\rm src}_{\rm start}}^{f^{\rm src}_{\rm end}} \frac{f'}{\dot{E}_{\rm GW} + \dot{E}_{\rm DF}} \frac{\dd E_{\rm orb} \left(f'\right)}{\dd f'} \dd f', \label{eq:Phi_total}
\end{align}
where $E_{\rm orb} = -\G \Mp \Ms/(2\aEMRI)$ is the orbital binding energy and 
\begin{align}
    \dot{E}_{\rm GW} &= -\frac{32\G^{7/3}}{5\cl^5} \eta^2 [\pi f (\Mps + \ME)]^{10/3}\,,  \label{eq:EdotGW} \\
    \dot{E}_{\rm DF} &= -4\pi \G^2 (\Mps + \ME)^2 \rho_{\rm DM}(\aEMRI) \eta^2 \frac{(1+v^2/\cl^2)^2}{1 - v^2/\cl^2}\frac{\ln \Lambda}{v}\,, \label{eq:EdotDF}
\end{align}
are the orbital energy losses due to gravitational waves \citep{Peters_1963} and dynamical friction with special relativistic corrections \citep{Chandrasekhar_1943, Mitra_2025}, respectively. Here $\eta = \Mps \ME/(\Mps + \ME)^2$ is the symmetric mass ratio, $\aEMRI = (\G(\Mps + \ME)/(\pi f)^2)^{1/3}$ is the orbital radius, $v = \pi \aEMRI f$ is the Newtonian orbital velocity of the EMRI, and $\Lambda = \aEMRI v^2/(\G \gamma_{\rm sp} \ME)$ with $\gamma_{\rm sp} = 7/3$ being the slope of the DM density profile \citep{Chandrasekhar_1943, Petts_2016}. One then obtains the dephasing, $\Delta \Phi$, by subtracting the phase accumulated in the presence of DM from that in vacuum, where $\dot{E}_{\rm DF} = 0$, over the common frequency interval $[f^{\rm src}_{\rm start}, f^{\rm src}_{\rm end}]$. We have neglected relativistic orbital corrections \citep{Karydas_2025}.
We note that our formula for dynamical friction assumes a static DM background. However, as can be seen in Figure~\ref{fig:rp_a_2D_hist_0_2}, most DM particles finish on highly eccentric weakly-bound orbits and will hence be moving at the same speed as the EMRI, if not faster. While previous work has shown that fast particles generally contribute sub-dominantly to semi-major axis evolution for steep profiles with an isotropic equilibrium velocity distribution \citep{Dosopoulou_2024}, this is a limitation of our work and our description serves as an approximation.

We determine $f^{\rm src}_{\rm end}$ by integrating Peters' equations \citep{Peters_1963, Peters_1964}, which describe the evolution of a binary system in a vacuum due to gravitational radiation reaction over a five-year observation window ($5/(1+z)$ years in the EMRI frame due to cosmological redshift) with initial GW frequency $f^{\rm src}_{\rm start}$. If the EMRI reaches a separation distance of $5 \, \Rschps$, then the run is terminated as we have not modelled smaller radii due to the breakdown of our 2.5PN description as well as spin effects becoming non-negligible. Otherwise the final frequency is given by
\begin{align}
    f^{\rm src}_{\rm end} = \left(\left(f^{\rm src}_{\rm start}\right)^{-8/3} - \frac{256 \pi^{8/3} \G^{5/3} \Mps \ME}{5 \cl^5 (\Mps + \ME)^{1/3}} \frac{\rm 5 \, yr}{1 + z}\right)^{-3/8}\,. \label{eq:f_end}
\end{align}
Additional energy losses due to dynamical friction will only speed up this inspiral, hence ensuring that $f^{\rm src}_{\rm end}$ will always be reached earlier than in a vacuum, i.e.,~before LISA's five-year observation window closes \citep{Colpi_2024}.

Additionally, in the limits $(f^{\rm src}_{\rm end} - f^{\rm src}_{\rm start})/f^{\rm src}_{\rm end} \ll 1$ and $\ME/\Mps \ll 1$, the leading order dependence on $\ME$ cancels, leaving only that in $\ln \Lambda$. However, this logarithmic scaling is very weak, making the dephasing independent of $\ME$ to leading order. 

Figure \ref{fig:Dephasing_z_mult_cell_mean} shows the observed dynamical-friction-induced dephasing experienced by a circular EMRI with secondary mass $\ME = 10 \, \Msolar$ after a single merger of mass ratio $q = \{$0.005, 0.03, 0.4$\}$ for MBH masses of $10^6 \, \Msolar$ (blue), $10^7 \, \Msolar$ (red), $10^8 \, \Msolar$ (green) and $10^9 \, \Msolar$ (black) and redshifts $z = 0$ and $z = 3$ as a function of the initial observed GW frequency, $f_{\rm start}^{\rm obs} = f^{\rm src}_{\rm start}/(1+z)$. We have shaded the region of the frequency spectrum that is observable by LISA in light grey, but for these MBH masses, the emitted GWs are often at lower frequencies. We have restricted the initial separation distance of the inspiralling EMRI to between $5\,\Rschps$ and $100\,\Rschps$, as this is the distance range that we consider in \S\ref{sec:Disruption of a Spikes by Individual Mergers}. For reference, we have highlighted the LISA detectability threshold, $\Delta \Phi = 1$.

It is clear that $q = 0.4$ mergers (dash-dotted line) and $q = 0.03$ mergers (dotted line) reduce the dephasing below the one-radian benchmark across the entire $10^6 \, \Msolar \leq \Mps \leq 10^9 \, \Msolar$ mass range at both redshifts. For $q = 0.005$ mergers, however, this is only true for EMRIs with initial semi-major axis $\aEMRIz \lesssim 19 \, \Rschps$ at $z = 0$ and $\aEMRIz \lesssim 85 \, \Rschps$ at $z = 3$.

We find that dephasing within the undepleted \GS-spike reaches a visible maximum at some frequency for $\Mps = 10^8\, \Msolar$ and $\Mps = 10^9\, \Msolar$. This maximum is located at the frequency at which $\dot{E}_{\rm DF} \approx \dot{E}_{\rm GW}$ -- the evolution is dominated by DF rather than GW emission at lower frequencies than it, and \emph{vice versa} for higher frequencies. Consequently, to the left of this maximum, the total number of cycles between $f_{\rm start}^{\rm obs}$ and $f_{\rm end}^{\rm obs}$ in the presence of DM approaches zero, while the number of cycles in vacuum, and hence dephasing, scales approximately linearly with $f_{\rm start}^{\rm obs}$. To the right, where GWs dominate the evolution, one can Taylor expand $1/(\dot{E}_{\rm DF} + \dot{E}_{\rm GW})$ in Eq.~\eqref{eq:Phi_total} to find that $\Delta \Phi$ which shrinks as the frequency increases due to the strong $\dot{E}_{\rm GW} \propto \aEMRI^{-5}$ scaling with $\aEMRI$, the EMRI's separation distance. Using the scalings present in Eqs.~\eqref{eq:EdotGW}, \eqref{eq:EdotDF}, and \eqref{eq:f_end}, one finds that $\Delta \Phi \propto \Mps^{-0.066}$ when $\dot{E}_{\rm GW} \gg \dot{E}_{\rm DF}$, $(f^{\rm src}_{\rm end} - f^{\rm src}_{\rm start})/f^{\rm src}_{\rm end} \ll 1$ and distance is measured in units of $\Rschps$, explaining why the dephasing appears almost independent of $\Mps$.

Finally, we also see a correlation with redshift. Specifically, systems at higher redshift exhibit less dephasing and have a lower observed GW frequency for the same EMRI separation distance. The former is explained by the fact that the observation time in the EMRI frame is $5/(1+z)$ yr while the latter is simply due to direct redshifting of the emitted GWs. Both effects would only become more pronounced at higher redshifts.

From Figure \ref{fig:Dephasing_z_mult_cell_mean} one identifies $q \approx 0.03$ as the mass ratio above which DM spikes no longer induce more than one radian of dephasing, making detection very difficult (we can see that the true minimum is somewhere within $0.005 \leq q \leq 0.03$). The next natural question to ask is how many MBHs have undergone such mergers, which we answer by turning to cosmological simulations. These will indicate how likely mergers of a given mass ratio are, and hence the MBH mass range over which we should expect MBHs to host or not host detectable spikes. We remark, however, that merger histories predicted by cosmological simulations can disagree and are quite uncertain \citep[e.g.,][]{Habouzit_2021, Habouzit_2022, Izquierdo_2026}. Since the results presented thus far are self-contained, they may be used independently to analyse spike detectability by treating the MBH merger histories as a free parameter, and combining them with MBH merger-trees from future cosmological simulations. 

\section{MBH Merger Histories}
\label{sec:MBH Merger Histories}
To estimate the BH merger histories of MBHs across cosmic time, we extract the evolving MBH population from the {\tt Horizon-AGN} simulation \citep{Dubois_2016}. Future gravitational wave detectors, such as LISA, are expected to detect EMRIs predominantly at redshifts below $z \sim $ 2\,--\,3 \citep{Babak_2017, Chapman_2025, Speri_2026}, whence we will focus on the merger history of MBHs from $z = 0$ to $z = 3$.

\subsection{MBH Mergers within Horizon-AGN}

{\tt Horizon-AGN} is a hydrodynamical cosmological simulation with a box length of $(100 \, h^{-1} \, {\rm Mpc})^3$ based on a WMAP-7 cosmology \citep{Komatsu_2011}, run down to $z=0$. It has a DM and stellar mass resolution of $8 \times 10^7 \, \Msolar$ and $2 \times 10^6 \, \Msolar$, respectively, and an adaptive resolution up to $\Delta x = 1 \, {\rm kpc}$ using a quasi-Lagrangian refinement scheme. In order to keep the minimum cell size approximately constant in physical units, a new maximum level of refinement is used every time the expansion scale factor doubles. {\tt Horizon-AGN} follows a standard complement of baryonic physics, including cooling down to $10^4 \, \rm K$, heating from a uniform UV background at $z_{\rm reion} = 10$, star formation and stellar feedback. For further details, see \citet{Dubois_2016}. 

Details of the model used to evolve MBHs and the properties of the resulting MBH population can be found in \citet{Volonteri_2016}. MBHs are placed with a seed mass of $10^5 \, \Msolar$ in cells where the gas and stellar densities exceed the threshold for star formation (hydrogen density of $n_0 = 0.1 \rm \, H \, cm^{-3}$), and the stellar velocity dispersion is larger than $100 \rm \, km \, s^{-1}$. MBHs are not allowed to form within 50 comoving kpc of an existing MBH. Accretion onto MBHs is modelled using a Bondi-Hoyle-Lyttleton prescription, with a density-based boost factor to account for unresolved dense gas following \citet{booth_2009}. Energy is returned to the gas in its vicinity at a rate of $\dot{E}_{\rm AGN} = \epsilon_f \epsilon_r \dot{M}_{\rm BH} \cl^2$ in one of two feedback modes: for inefficient accretion (Eddington ratio $f_{\rm Edd} < 0.01$), energy is returned to the gas using a bipolar outflow with a velocity of $10^4 \rm \, km \, s^{-1}$. For efficient accretion (Eddington ratio $f_{\rm Edd} \geq 0.01$), feedback energy is deposited as thermal energy, using a coupling efficiency of $\epsilon_f = 0.15$. The radiative efficiency of the MBH is $\epsilon_r = 0.1$ and accretion is capped at the Eddington limit.

\begin{figure}
    \centering
    \includegraphics[width=0.49\textwidth]{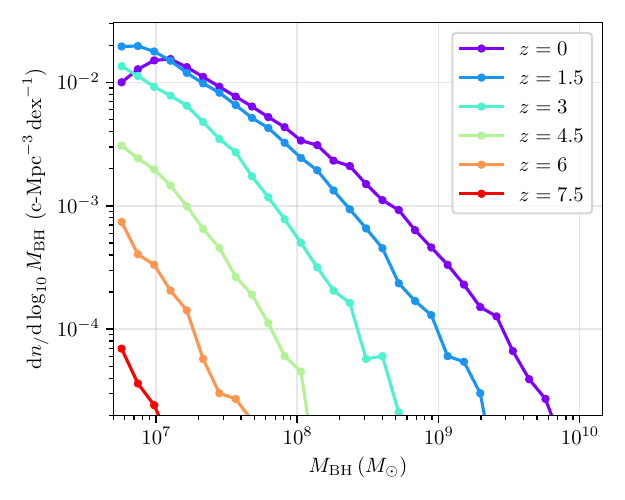}
     \caption{
     The MBH mass function per cubic comoving Mpc (c-Mpc) per logarithmic MBH mass interval at redshifts $z = \{$0, 1.5, 3, 4.5, 6, 7.5$\}$ predicted by {\tt Horizon-AGN}.
     }
    \label{fig:mass_dist_for_diff_zs}
\end{figure}

\begin{figure*}
    \centering
    \includegraphics[width=0.99\textwidth]{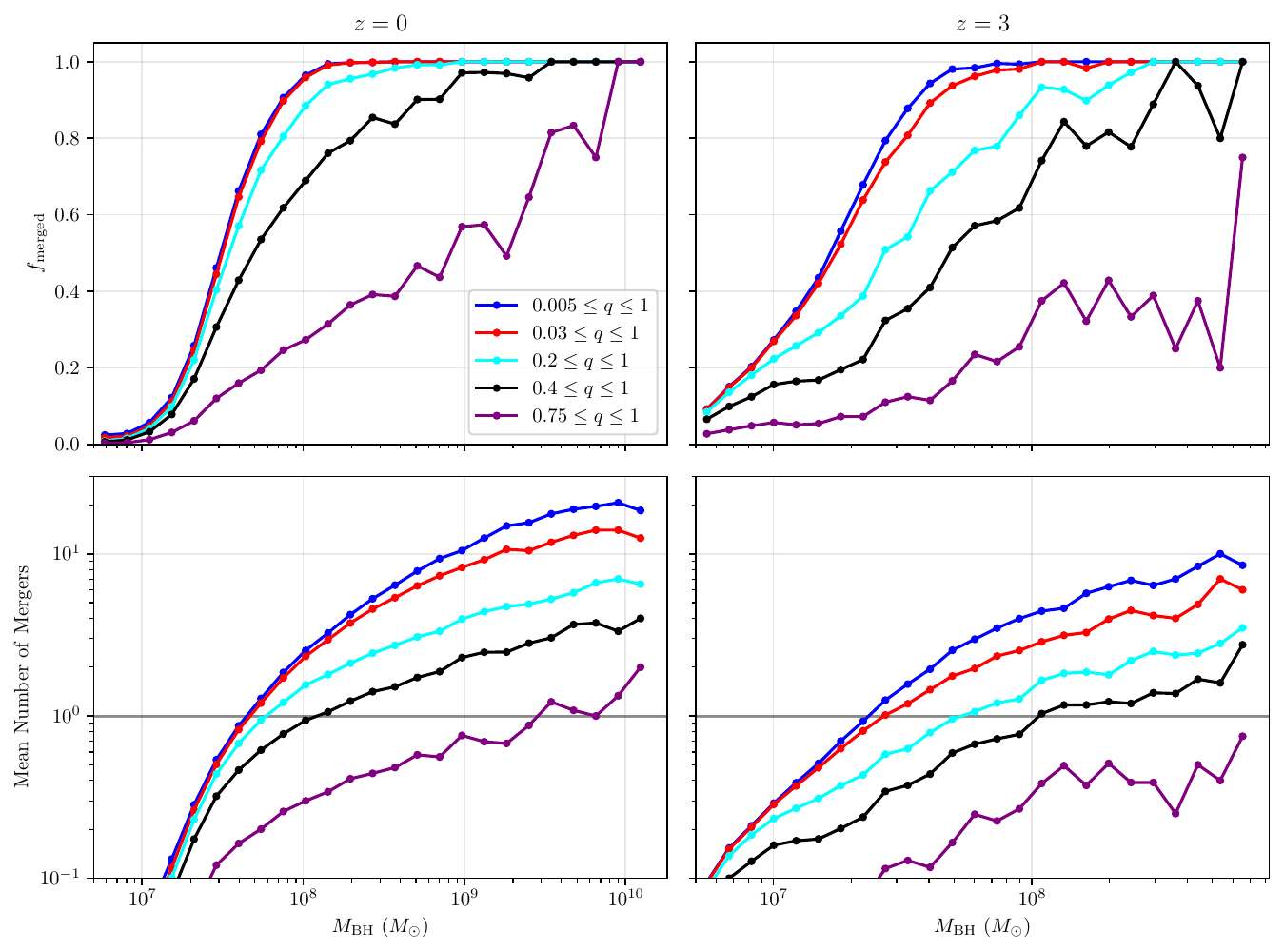}
     \caption{
     The fraction, $f_{\rm merged}$, of MBHs that have undergone at least one merger (top row) and the average number of mergers per MBH (bottom row) for various mass-ratio thresholds at $z = 0$ (left column) and $z = 3$ (right column) predicted by {\tt Horizon-AGN}.
     }
    \label{fig:merger_fractions_and_ave_num_mergers_4_panels}
\end{figure*}

\begin{figure*}
    \centering
    \includegraphics[width=0.99\textwidth]{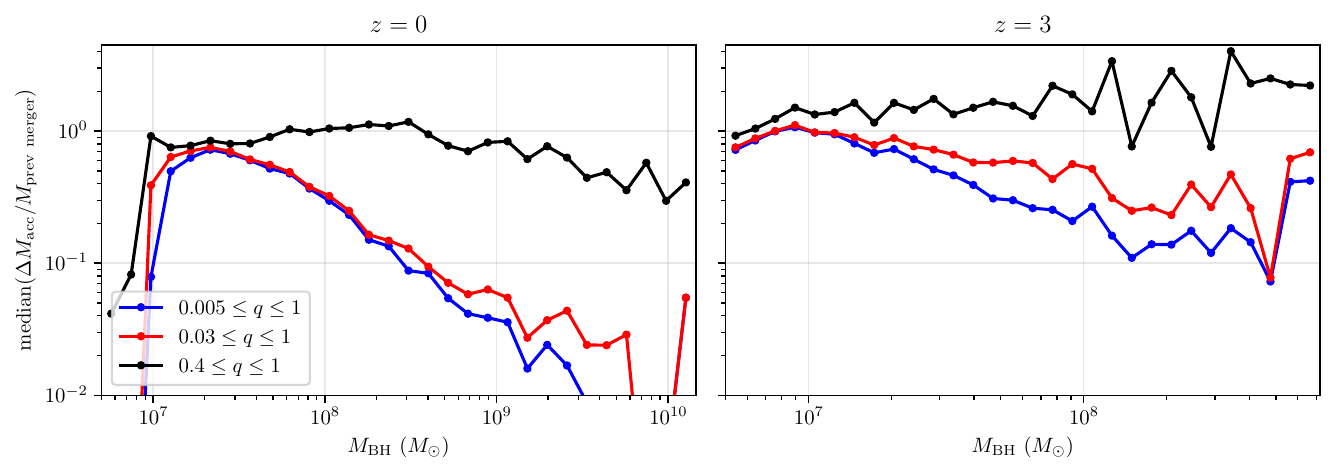}
     \caption{
     The median fractional increase in an MBH's mass due to accretion since its last merger with mass ratio $q \geq \{$0.005, 0.03, 0.4$\}$ as a function of mass at $z = 0$ and $z = 3$. See text for details. 
     }
    \label{fig:median_fractional_mass_increase_vs_current_mass_z_0_3}
\end{figure*}

In {\tt Horizon-AGN}, MBHs are allowed to move in the potential of their host galaxy and their dynamics are explicitly followed. To compensate for the unresolved dynamical friction between an MBH and the background gas, which allows MBHs to lose angular momentum and settle in the centres of their host galaxy, an explicit drag force $F_{\rm drag} = f_{\rm gas} 4 \pi \nu \rho (\G \mMBH/ c_{s})^2$ is added to particles, following \citet{ostriker_1999}. Here $f_{\rm gas}$ is a factor between 0 and 2, $\nu$ is the same density-dependent boost used for accretion, and $\rho$ and $c_{\rm s}$ are the density and gas sound speed at the location of the MBH, respectively. 

Furthermore, two MBHs are merged if their separation distance $d$ is less than $4 \, \Delta x$ ($\sim 4 \rm \, kpc$), where $\Delta x$ is a cell size, and they are energetically bound in vacuum, i.e., the relative velocity is less than $\sqrt{2\G\Mps/d}$. We identified MBH mergers using information on the fine integration time-step of the simulation. When two MBHs merge, the lighter (secondary) MBH is removed from the simulation while the heavier (primary) MBH is kept and its properties are updated to represent the merger remnant. Thus, for any given MBH within the simulation, we can obtain the number of mergers it has undergone and the mass ratios of these mergers.

Because the DM particle mass in {\tt Horizon-AGN} is $8 \times 10^7 \, \Msolar$, MBHs at the lower end of the MBH mass spectrum do not have well-resolved environments as they often reside in smaller galaxies. Further, the MBH seed mass is $10^5 \, \Msolar$. This restricts the range of validity of the MBH mass spectrum to $\mMBH \gtrsim 5 \times 10^6 \, \Msolar$ -- below this, the mass spectrum is incomplete and merger histories cannot be reliably deduced \citep{Volonteri_2016}. We note that we treat mergers as instantaneous when the simulation merger criteria are met and do not account for the delay introduced by binary evolution \citep{volonteri_2020}, which could change the redshift at which mergers occur. However, even for a delay of, say, 1 Gyr, our conclusions are mostly unchanged, particularly at low redshift.

\subsection{Results}

Figure \ref{fig:mass_dist_for_diff_zs} shows the MBH mass function per cubic comoving Mpc (c-Mpc) per logarithmic mass interval for $z = \{$0, 1.5, 3, 4.5, 6, 7.5$\}$ for masses above $5 \times 10^6 \, \Msolar$. The dip in the number of MBHs with $\mMBH \lesssim 10^7 \, \Msolar$ at $z = 0$ is an example of an unphysical artefact that results from the finite mass resolution and finite seed mass \citep{Volonteri_2016}. As expected, the distribution broadens towards higher masses at lower redshifts -- this is due to accretion and MBH mergers gradually growing the range of accessible masses.

Figure \ref{fig:merger_fractions_and_ave_num_mergers_4_panels} shows the statistics of MBH mergers of mass ratios $q_{\rm min} \leq q \leq 1$, where $q_{\rm min} = \{$0.005, 0.03, 0.2, 0.4, 0.75$\}$, as a function of MBH mass for $z = 0$ (left column) and $z = 3$ (right column). The top row shows the fraction, $f_{\rm merged}$, of MBHs that have undergone at least one merger in their lifetime, while the bottom row shows the average number of mergers that have occurred during their lifetime. Note the different mass ranges for $z = 0$ and $z = 3$.

While mergers with mass ratios close to unity are relatively rare, smaller mergers with mass ratios down to $q = 0.005$ are much more common. For $q \geq 0.03$, the merger fraction at $z = 0$ is approximately $\sim 50\%$ at $\mMBH = 3 \times 10^7 \, \Msolar$, $\sim 95\%$ at $\mMBH = 10^8 \, \Msolar$, and reaches 100\% by $\mMBH = 2 \times 10^8 \, \Msolar$. Such mergers are even more frequent at $z = 3$, where these thresholds instead sit at $\mMBH = 2 \times 10^7 \, \Msolar$, $\mMBH = 5 \times 10^7 \, \Msolar$, and $\mMBH = 10^8 \, \Msolar$, respectively. The average number of these smaller mergers per MBH exceeds unity at $\mMBH \sim 5 \times 10^7 \, \Msolar$ for $z = 0$ and $\mMBH \sim 2.5 \times 10^7 \, \Msolar$ for $z = 3$. 

The merger fraction and the average number of mergers, for a given mass, actually decrease as we move from $z = 3$ to $z = 0$. At high redshift, there has been little time for accretion to increase MBH masses significantly. Consequently, the most massive black holes must have obtained a significant portion of their mass through repeated mergers, causing a bias towards hierarchical merger formation. Meanwhile, black holes at $z = 0$ have had much more time to grow via accretion and, as a result, many MBHs can reach the same mass while experiencing fewer mergers.

Different cosmological simulations predict different MBH population statistics \citep{Habouzit_2021, Habouzit_2022, Izquierdo_2026}. Thus, while emphasising that {\tt Horizon-AGN} is well suited for MBH modelling, we attempted to find the mass distributions and MBH merger histories predicted by {\tt SIMBA} \citep{Dave_2019},\footnote{The {\tt SIMBA} simulation data are publicly available -- \url{https://simba.roe.ac.uk/}.} {\tt IllustrisTNG300} \citep{Nelson_2018},\footnote{The {\tt IllustrisTNG300} simulation data are publicly available -- \url{https://www.tng-project.org/data/}.} and {\tt FLAMINGO} \citep{Schaye_2023}.\footnote{The {\tt FLAMINGO} data can be found at their webpage -- \url{https://flamingo.strw.leidenuniv.nl}.} Unfortunately, however, in the case of {\tt SIMBA} and {\tt IllustrisTNG300}, the data files contain spurious points where BHs are spontaneously removed from the simulation without a corresponding merger. The underlying cause for this likely relates to errors in output files which often coincide with simulation restarts \citep{Izquierdo_2026}. Resolving these issues would require analysing snapshots to cross-check merger events against host galaxy catalogues. This is beyond the scope of our work and further motivates the use of {\tt Horizon-AGN}, which does not face such issues.  Further, {\tt FLAMINGO} had insufficient resolution of merger mass ratios.

Combining these merger histories with our dephasing calculations brings us to an important conclusion: MBH-MBH mergers provide a strong upper limit on the MBH mass range over which collisionless DM spikes are likely to be detectable. {\tt Horizon-AGN} predicts that, depending on redshift, mergers with mass ratio $q \gtrsim 0.03$ -- individually capable of suppressing the DM-induced EMRI dephasing to below one radian -- become very common and effectively guaranteed for $\mMBH \gtrsim 5 \times 10^7 \, \Msolar$ for $z = 3$ and $\mMBH \gtrsim 10^8 \, \Msolar$ for $z = 0$. Furthermore, many MBHs in this mass range undergo several such mergers, making our single-merger depletion estimates highly conservative as repeated mergers would erode the spike even further.

\section{Discussion}
\label{sec:Discussion}

\subsection{Re-formation of the Spike During Adiabatic MBH Growth}

\begin{figure*}
    \centering
    \includegraphics[width=0.99\textwidth]{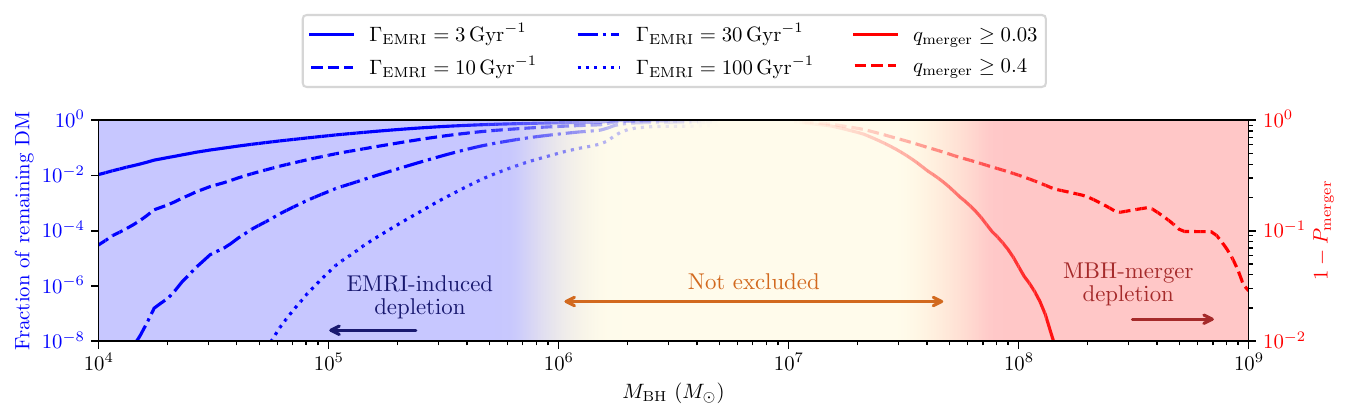}
     \caption{
     The range of MBH masses that have not been excluded as suitable DM spike hosts by either EMRI-induced or MBH-merger-induced DM depletion. At low masses, we show the fraction of DM density remaining at 10 Schwarzschild radii by $z = 3$, as a function of MBH mass and EMRI rate, $\Gamma_{\rm EMRI}$. At higher masses, the MBH-merger depletion shows the probability that no MBH merger has occurred by $z = 0$ as a function of mass and the merger's mass ratio, which we have labelled $q_{\rm merger}$ for clarity. $P_{\rm merger}$ is the probability that a merger has occurred.
     }
    \label{fig:Goldilocks_zone}
\end{figure*}

A remaining question is whether DM spikes can re-form through accretion-driven MBH growth between successive mergers. {\tt Horizon-AGN}, however, predicts insufficient accretion-driven growth for this process to be appreciable. During adiabatic MBH growth, the product of a DM particle's semi-major axis and the MBH mass is conserved \citep{Gondolo_1999}. Rebuilding a spike across the full $5\,\Rschps$--$100\,\Rschps$ range considered here would therefore require the MBH mass to increase by a factor of $ \gtrsim \sqrt{20} \approx 4.5$ (we keep distance measured in Schwarzschild radii). This would also strongly depend on the extent of the spike's depletion at radii larger than $100\,\Rschps$, which we have not modelled.

Figure~\ref{fig:median_fractional_mass_increase_vs_current_mass_z_0_3} shows the median fractional MBH mass growth through accretion since the most recent merger, $\Delta M_{\rm acc}/M_{\rm prev\,merger}$, predicted by {\tt Horizon-AGN} as a function of MBH mass at $z=3$ and $z=0$, for mergers with mass ratio $q \geq \{$0.005, 0.03, 0.4$\}$. Here, $\Delta M_{\rm acc}$ is the mass accreted since the most recent merger, and $M_{\rm prev\,merger}$ is the MBH mass immediately after that merger. Although MBHs can grow substantially following mergers with $q \geq 0.4$, the frequent occurrence of mergers with $q \geq \{0.005, 0.03\}$ limits the typical intervening fractional increase in mass to $\mathcal{O}(1)$, and in most cases to $\ll 1$. This behaviour is consistent with previous results \citep{Dubois_2014, Pacucci_2020, Fang_2025}. We therefore expect merger-depleted DM spikes to undergo negligible re-formation between mergers with mass ratio $q \geq 0.03$.\footnote{Naturally this is under {\tt Horizon-AGN}'s accretion prescription, which does not include super-Eddington accretion. Developing a more general model is beyond the scope of this work.}

\subsection{Detectability Mass Window}
\label{subsec:Detectability Mass Window}

We have found that a single MBH merger with mass ratio $q \gtrsim 0.03$ is sufficient to deplete DM spikes to 0.1\% of the original density, and often much less, at distances between $5 \, \Rschps$ and $100 \, \Rschps$ from the MBH, bringing gravitational-wave dephasing to below the one-radian benchmark across all masses and redshifts considered. Even a lighter merger with $q = 0.005$ is still capable of reducing detection prospects significantly. Predictions from {\tt Horizon-AGN} showed that, depending on redshift, mergers with mass ratio $q \geq 0.03$ are almost guaranteed ($\gtrsim 95\%$ occurrence) for MBHs of mass $5 \times 10^7 \, \Msolar$--$10^8 \, \Msolar$. Consequently, MBHs heavier than $5 \times 10^7 \, \Msolar$ are unlikely hosts for DM spikes, particularly at higher redshifts. This also suggests that the strength of the imprints DM spikes are expected to have on the stochastic GW background \citep{Ghoshal_2023, Alonso_Alvarez_2024, Hu_2025, Chen_2025, Shen_2025b, Shen_2025a, Tiruvaskar_2026} may require re-evaluation.

Our previous work, which focused on EMRI-induced DM depletion \citep{Sharpe_2026a, Sharpe_2026b}, found that typical predicted EMRI rates of $100$--$300 \, \Gyr^{-1}$ induce reductions in densities sufficient to prevent detectability with LISA for MBHs of mass less than $\mMBH \sim 10^6 \, \Msolar$ by $z = 3$. The depletion becomes even more substantial for lower redshifts. This, in combination with the work presented here, significantly narrows the mass range of MBHs that are expected to host DM spikes capable of inducing more than one radian of dephasing to $10^6 \, \Msolar \lesssim \mMBH \lesssim 5 \times 10^7 \, \Msolar$. Figure \ref{fig:Goldilocks_zone} shows this mass window and displays the fraction of remaining DM after EMRI-induced depletion for various EMRI rates in blue, and the probability that no MBH merger with mass ratio greater than $q$ has occurred for $q = \{0.03, 0.4\}$ in red. We show conservative estimates here by selecting $z = 3$ for EMRI-induced depletion, as younger systems have had less time to be depleted, and $z = 0$ for MBH mergers, since lower-redshift MBHs are less affected by mergers, for a given MBH mass. The transition between the two regimes is blurred to represent the non-trivial redshift dependence, which we now describe.

\subsection{Redshift Dependence of Suitable Spike Hosts}

{\tt Horizon-AGN} merger-history predictions indicate that black holes at higher redshifts are actually less promising candidates for hosts of detectable DM spikes. This is due to an assembly effect -- young black holes have not had much time to gain mass via accretion and hence the heaviest ones must have obtained a significant portion of their mass via repeated hierarchical mergers, naturally increasing the average number of mergers. Figure~\ref{fig:merger_fractions_and_ave_num_mergers_4_panels} illustrates this: a $10^7 \, \Msolar$ MBH, for example, at $z = 3$ will, on average, have undergone more mergers than one at $z = 0$. Moreover, higher redshifts provide lower observed GW frequencies, for a given emission frequency, due to cosmological redshift (see Fig.~\ref{fig:Dephasing_z_mult_cell_mean}). Thus, this $5 \times 10^7 \, \Msolar$ upper bound is slightly lower at higher redshifts. For LISA in particular, which is primarily sensitive to EMRIs around MBHs of mass $\lesssim 10^{7} \, \Msolar$, DM spikes are more likely to be detected at \textit{lower} redshifts.

\subsection{Modelling Assumptions}
\label{subsec:Model Asses}

Finally, we summarise the assumptions underlying our work.

\begin{itemize}
    \item Our one-radian reference benchmark is motivated by LISA's approximate one-radian-dephasing detection threshold. Most EMRI systems we consider, however, sit below LISA's frequency band and are better source candidates for other future GW detectors. These include ASTROD-GW \citep{Ni_2013}, $\mu$Ares \citep{Sesana_2021, Perego_2026}, or lunar/satellite laser ranging interferometers \citep[see, e.g.,][ and references therein]{Fedderke_2022, Foster_2025}. While the exact dephasing detectability threshold depends on many specifics, the results presented here can be directly used for understanding future detection opportunities.

    \item The distribution of MBHs with mass less than $5 \times 10^6 \Msolar$ is unknown due to {\tt Horizon-AGN}'s finite mass resolution. While we have not considered the merger history of such MBHs, we have used them as secondaries for mergers. Besides, the MBH seed mass is $10^5 \, \Msolar$, meaning that the minimum mass ratio {\tt Horizon-AGN} allows for an MBH of mass $\mMBH$ is $10^5 \, \Msolar/\mMBH$. While a self-consistent treatment is desirable, this restricted range of validity is an unavoidable caveat of cosmological simulations and their finite mass resolutions. This does not, however, affect our conclusions on the depletion induced by an individual merger in \S\ref{sec:Disruption of a Spikes by Individual Mergers} and its effect on GW dephasing in \S\ref{sec:Detectability of DM Spikes}.

    \item We have assumed the secondary BH does not host its own DM spike. Integrating Eq.~\eqref{eq:rho_DM} from $r = 0$ to $r = r_{\rm sp}$ and using $r_{\rm sp} \propto r_{\rm SoI} \propto \mMBH^{0.543}$ shows that the total spike mass scales as $\mMBH^{1.629}$. Assuming the DM orbiting the secondary would undergo similar depletion to that orbiting the primary and be left with a similar distribution, the total post-merger DM densities would only increase by $\sim 22\%$ for a heavy $q = 0.4$ merger, and $\sim 0.3\%$ for a lighter $q = 0.03$ merger. When considered together with the reductions by multiple orders of magnitude induced by the merger itself, this addition would leave us with qualitatively identical results.
    
    \item One fundamental assumption implicit in {\tt Horizon-AGN}'s MBH-merging prescription is that mergers occur instantaneously once the 4 kpc separation condition is realised. This is due to length-scales smaller than this being unresolved. 
    Accounting for dynamical friction, binary hardening and GW-inspiral timescales may have an effect on merger rates \citep{Haiman_2009,Ellis_2024}. However, our results are mostly insensitive to any delay of $\mathcal{O}(1)\,$Gyr, particularly at low redshift.

    \item We have assumed that both MBHs involved are initially non-spinning. It has been shown that spin is only important within the central $\sim 5$ Schwarzschild radii \citep{Mitra_2025} -- distances which we do not consider. For this reason, as well as the fact that escape velocities are almost completely unchanged (see Appendix \ref{appsec:Escape velocities of (non-)rotating MBHs}), we have taken the remnant MBH as non-spinning. However, a more accurate model would include spin effects.

    \item We have run a finite number of DM--MBH--MBH three-body simulations, leading to uncertainties in the final density profiles. This is particularly true for depletions greater than $\sim 6$ orders of magnitude. However, in such instances, the depleted densities put the DM-induced dephasing well below our one-radian benchmark.

    \item We have ignored any GW-induced recoil of the merger remnant. It has been shown that the maximum recoil velocity from a merger of initially non-spinning black holes, which we assume here, is about $175 \rm \, km \, s^{-1}$ \citep{Gonz_lez_2007a}. This increases to $\sim 5000 \rm \, km \, s^{-1}$ for rotating MBHs with partially aligned spins \citep{Lousto_2011}. This can be larger than the apoapsis velocity of the weakly bound orbits occupied by most post-merger survivors, suggesting it could have an important effect on the post-merger density profile. This remains a limitation of our work.

    \item We have assumed the DM is collisionless. Self-interacting DM \citep{Tiruvaskar_2026}, self-annihilating DM \citep{Ullio_2001}, fuzzy DM \citep[e.g.,][]{Davies_2020,Aurrekoetxea_2024b,Aurrekoetxea_2024a}, and bosonic DM \citep{Brito_2020} models all predict significantly different behaviour regarding both spike formation and spike evolution. Exploring such systems is a future direction of this work.
\end{itemize}

\section{Conclusion}
\label{sec:Conclusion}
Collisionless DM spikes around MBHs can be significantly depleted by individual merger events. We have found that any merger with a mass ratio as low as $q = 0.03$ reduces a DM spike's density by roughly 3 orders of magnitude at 100 Schwarzschild radii and about 5 orders of magnitude at 5 Schwarzschild radii across all polar angles relative to the binary's plane for any MBH primary mass. Larger mass ratios induce even greater depletion. This causes DM-induced GW dephasings to fall below the one-radian dephasing detectability benchmark for MBHs of mass $10^6 \, \Msolar \leq M_{\rm BH} \leq 10^9 \, \Msolar$ and all semi-major axes between 5 and 100 MBH Schwarzschild radii for redshifts $z \leq 3$. In fact, detection prospects are significantly reduced by mergers with mass ratios all the way down to $q = 0.005$.

We connect these results to an MBH's cosmic history using {\tt Horizon-AGN} -- a cosmological simulation specifically designed to improve the tracking of MBH formation and evolution. {\tt Horizon-AGN} predicts that the fraction of MBHs that have undergone a merger with mass ratio $q \geq 0.03$ increases with MBH mass. In particular, at $z = 0$, such mergers have occurred for $\sim 50$\% of MBHs with masses $3 \times 10^7 \, \Msolar$ and 95\% of MBHs with masses $10^8 \, \Msolar$. At $z = 3$, we expect that 95\% of MBHs with masses $5 \times 10^7 \, \Msolar$ have undergone such a merger, reaching 100\% by $10^8 \, \Msolar$. MBHs at lower redshifts are better candidates for hosting detectable DM spikes because, at fixed mass, the fraction of MBHs that have undergone a merger increases with redshift. Further, accretion rates predicted by {\tt Horizon-AGN} are too low to allow accretion-driven adiabatic MBH growth between mergers to cause a spike to regrow after it has been depleted by a merger.

This work, along with our EMRI-induced depletion studies \citep{Sharpe_2026a, Sharpe_2026b}, suggests that DM spikes around MBHs across a broad mass range are unlikely to induce more than one radian of dephasing and leaves just $10^6 \, \Msolar \lesssim \mMBH \lesssim 5 \times 10^{7} \,\Msolar$ -- we emphasise that this upper limit is bottlenecked by the finite mass resolution of cosmological simulations. Interestingly, Sgr A$^*$ lies within this preferred mass range and may therefore be a promising candidate. While the exact boundaries of this window depend on various parameters such as assumed EMRI rates and the MBH's redshift, we have demonstrated that black hole mergers deplete DM spikes substantially.

\section*{Acknowledgements}
\label{Acknowledgements}

C.S.~acknowledges funding from an STFC studentship. T.F.M.S.~acknowledges support from a Royal Society University Research Fellowship (URF-R1-231065), the Friends of the Institute for Advanced Study Fund and the Simons Foundation through the Black Holes and Strong Gravity program with Award No.~SFI-MPS-BH-00012593-10. Y.B.G.'s research is supported by the Royal Commission for the Exhibition of 1851.~R.S.B.~acknowledges support from a UKRI Future Leaders Fellowship (grant code: MR/Y015517/1). B.K.~acknowledges support from the Hungarian Academy of Sciences (MTA) through the MTA Distinguished Guest Scientist Fellowship Programme 2026. C.S.~acknowledges the use of OpenAI's {\tt Codex} to assist with code development and the production of plots.

\section*{Data Availability}
\label{Data Availability}

The {\tt Horizon-AGN} simulation is described in \citet{Dubois_2016} and on the project website.\footnote{https://www.horizon-simulation.org/} 
Any other data will be made available upon reasonable request to the corresponding author. \multistar documentation is available online.\footnote{\url{https://2sn.erc.monash.edu/multistar/doc/index.html}.}



\bibliographystyle{mnras}
\bibliography{example} 




\appendix

\section{Mass Independence of Orbital Evolution}
\label{appsec:Over-all Mass Independence of Orbital Evolution}

Here, we show that the orbital evolution of an MBH binary is independent of the individual masses of the components, and depends only on their mass ratio, if distances and time are measured in units of $\Rschps$ and $\Rschps/\cl$, respectively. Firstly, there are only 7 length scales in our three-body system: $\Rschp$, $\Rschs$, $\aMBHB$, and $x^\alpha$, with $x^\alpha = (ct, \V{x})$ being the spacetime coordinate of the DM particle. Consequently, the dimensionless metric tensor, $g_{\mu \nu}$ which describes this system as well as the geodesic path of the DM particle, can only depend on ratios of these 7 parameters. That is, 
\begin{align}
    g_{\alpha \beta} = g_{\alpha \beta}\left(q, \frac{\aMBHB}{\Rschps}, \frac{x^\mu}{\Rschps} \right) \,,
\end{align}
where $\mu \in \{0,1,2,3\}$. 

Let us now define the transformation $x^\alpha \to \tilde{x}^{\alpha} = x^\alpha/\Rschps$. The metric in these new coordinates is given by 
\begin{align}
    \tilde{g}_{\mu \nu}\left(q, \tilde{\aMBHB}, \tilde{x}^\mu\right) = \Rschps^2 g_{\mu \nu}\left(q, \tilde{\aMBHB}, \tilde{x}^\mu\right)\,,
\end{align}
where $\tilde{\aMBHB} = \aMBHB/\Rschps$. When computing the Christoffel symbols, any overall $\tilde{x}^{\mu}$-independent factors, namely $\Rschps^2$ in this case, cancel out and the geodesic equation only depends on $\tilde{x}^{\mu}$. Additionally, invariance of the Christoffel symbols implies invariance of the Ricci tensor, $R_{\mu \nu}$, leaving the vacuum Einstein field equations unchanged -- here we are ignoring the cosmological constant as well as the DM particle's contribution to the stress energy tensor. Thus, the evolution of the DM particle as well as the binary depends on the component masses only through $q$, at fixed dimensionless initial conditions, as long as time and distance are measured in units of $\Rschps/\cl$ and $\Rschps$, respectively.

\section{Escape velocities of (non-)rotating MBHs}
\label{appsec:Escape velocities of (non-)rotating MBHs}

Here, we compare the escape velocities of rotating (Kerr) and non-rotating (Schwarzschild) black holes and show that the spin corrections to the escape velocities are negligible, justifying the assumption of the post-merger remnant MBH being non-spinning.

We begin by calculating the dimensionless specific energy of a DM particle around a rotating MBH with spin parameter $a$, defined as the conserved quantity $E = -u_\mu \xi^\mu$ associated with the Kerr metric's Killing vector $\xi^\mu = (1,0,0,0)$ where $u_\mu$ is the covariant 4-velocity. Consider a DM particle orbiting an MBH with mass $\mMBH$ and angular momentum $J$. We work with the following parameters and dimensionless coordinates
\begin{align}
    \begin{aligned}
        \Rg &= \frac{\G \mMBH}{\cl^2}, 
        & a &= \frac{J}{c \Rg \mMBH}, 
        & \tilde{t} &= \frac{ct}{\Rg}, 
        & \tilde{\tau} &= \frac{c\tau}{\Rg}, \\
        \tilde{r} &= \frac{r}{\Rg}, 
        & \tilde{v}_r &= \frac{\dd \tilde{r}}{\dd \tilde{t}}, 
        & \tilde{v}_\theta &= \tilde{r} \frac{\dd \theta}{\dd \tilde{t}}, 
        & \tilde{v}_\phi &= \tilde{r} \sin \theta \frac{\dd \phi}{\dd \tilde{t}}\,,
    \end{aligned}
\end{align}
where $t, \tau, r, \theta$, and $\phi$ are the coordinate time normalised at infinity, the proper time, the radial coordinate, the polar angle, and the azimuthal angle, respectively -- this is the standard Boyer–Lindquist convention. Using these coordinates ensures the results discussed here are independent of mass, which follows from applying $q = 0$ to the proof presented in Appendix~\ref{appsec:Over-all Mass Independence of Orbital Evolution}. We further define the useful parameters
\begin{align}
    \Sigma = \tilde{r}^2 + a^2 \cos^2 \theta, \quad \Delta = \tilde{r}^2 - 2 \tilde{r} + a^2, \quad F = 1 - \frac{2\tilde{r}}{\Sigma} \,.
\end{align}
In these coordinates, the proper time is given by
\begin{align}
    \begin{aligned}
        -\dd \tilde{\tau}^2 = - F \, \dd \tilde{t}^2 &- \frac{4a\tilde{r} \sin^2 \theta}{\Sigma} \, \dd \tilde{t} \, \dd \phi + \frac{\Sigma}{\Delta} \, \dd \tilde{r}^2 + \Sigma \, \dd \theta^2 \\
        &+ \left(\tilde{r}^2 + a^2 + \frac{2a^2 \tilde{r} \sin^2 \theta}{\Sigma} \right) \sin^2 \theta\,  \dd \phi^2\,.
    \end{aligned}
\end{align}
Rearranging gives
\begin{align}
    \begin{aligned}
        u^{\tilde{t}} = \left[ \vphantom{\frac{1}{1}}\right.F &+ \frac{4 a \sin \theta}{\Sigma} \tilde{v}_\phi - \frac{\Sigma}{\Delta} \tilde{v}_r^2 - \frac{\Sigma}{\tilde{r}^2}\tilde{v}_\theta^2
        \\ & \qquad \qquad - \left(1 + \frac{a^2}{\tilde{r}^2} + \frac{2a^2 \sin^2 \theta}{\tilde{r} \Sigma} \right) \tilde{v}_\phi^2\left. \vphantom{\frac{1}{1}}\right]^{-1/2}\,.
    \end{aligned}
\end{align}
The conserved quantity is then
\begin{align}
    E = -u_{\tilde{t}} = - u^{\mu }g_{\mu \tilde{t}} = \left(F + \frac{2a \sin \theta}{\Sigma} \tilde{v}_\phi \right) u^{\tilde{t}}\,.
\end{align}
By setting $E = 1$, specifying a position and velocity direction, and rearranging for the speed, one obtains the minimum velocity magnitude required for escape.

We find that even for a maximally spinning Kerr black hole ($a = 1$), the escape velocity will be within 1.6\% of that for a non-rotating MBH, across all positions with radii $\geq 5\Rschps$ and all outward velocities. Any spins less than unity give even closer agreement. Consequently, MBH spin has a negligible effect on the ejection fraction of DM particles, justifying our Schwarzschild approximation.

\section{Initial MBH Binary Semi-major Axes}
\label{appsec:Initial MBH Binary Semi-major Axes}

Here, we state the initial MBH binary semi-major axes used in our simulations for reproducibility purposes. We varied them as a function of DM periapsis, DM eccentricity, and binary mass ratio. For the ordered list $\rpDMz/\Rschps = \{$5, 10, 20, 40, 60, 100$\}$, we chose
\begin{align}
    \frac{\aMBHBz}{\Rschps} =
    \begin{cases}
        \{15,25,40,70,100,100\}, & \eDMz=0.1,\ \ 0\le q\le 0.01,\\
        \{20,40,50,70,100,100\}, & \eDMz=0.1,\ \ 0.01<q\le 0.2,\\
        \{25,40,80,100,100,100\}, & \eDMz=0.1,\ \ q>0.2,\\
        \{20,40,50,100,100,100\}, & \eDMz=0.5,\ \ 0\le q\le 0.01,\\
        \{40,60,80,100,100,100\}, & \eDMz=0.5,\ \ 0.01<q\le 0.2,\\
        \{50,80,100,100,100,100\}, & \eDMz=0.5,\ \ q>0.2,\\
        \{100,100,100,100, & \multirow{2}{*}{$\eDMz=0.9\,.$}\\
        \phantom{\{}100,100\}, & \\
    \end{cases}
\end{align}
While these choices are for computational efficiency and are not physically motivated, we confirmed that our results are independent of the exact values.

\bsp	
\label{lastpage}
\end{document}